\documentclass[aps,pra,twocolumn,superscriptaddress,showpacs]{revtex4-1}
\usepackage[version=3]{mhchem}
\usepackage[english]{babel}
\usepackage{dcolumn}
\usepackage{graphicx}
\usepackage{amsmath}
\usepackage{longtable}
\usepackage{bigstrut}
\usepackage{mathtools}
\usepackage{tabularx}
\usepackage{multirow}
\allowdisplaybreaks[1]
\begin{document}

\title{Sensitivity to a possible variation of the Proton-to-Electron Mass Ratio of Torsion-Wagging-Rotation 
Transitions in Methylamine (\ce{CH_{3}NH_{2}})}
\author{Vadim V. Ilyushin}
\affiliation{Institute of Radio Astronomy of NASU, Chervonopraporna 4, 61002 Kharkov, Ukraine}
\author{Paul Jansen}
\affiliation{Institute for Lasers, Life and Biophotonics, VU University Amsterdam, 
De Boelelaan 1081, 1081 HV Amsterdam, The Netherlands}
\author{Mikhail G. Kozlov}
\affiliation{Petersburg Nuclear Physics Institute, Gatchina 188300, Russia}
\author{Sergei A. Levshakov}
\affiliation{Ioffe Physical-Technical Institute, 194021 St. Petersburg, Russia}
\author{Isabelle Kleiner}
\affiliation{Laboratoire Interuniversitaire des Syst\`{e}mes Atmosph\'{e}riques (LISA), CNRS UMR 7583 et Universit\'{e}s Paris 7 et Paris Est, 61 avenue du G\'{e}n\'{e}ral de Gaulle, 94010 Cr\'{e}teil C\'{e}dex, France}
\author{Wim Ubachs}
\affiliation{Institute for Lasers, Life and Biophotonics, VU University Amsterdam, 
De Boelelaan 1081, 1081 HV Amsterdam, The Netherlands}
\author{Hendrick L. Bethlem}
\affiliation{Institute for Lasers, Life and Biophotonics, VU University Amsterdam, 
De Boelelaan 1081, 1081 HV Amsterdam, The Netherlands}

\date{\today}

\begin{abstract}
We determine the sensitivity to a possible variation of the proton-to-electron mass ratio $\mu$
for torsion-wagging-rotation transitions in the ground state of methylamine (\ce{CH_{3}NH_{2}}).
Our calculation uses an effective Hamiltonian based on a high-barrier tunneling formalism combined 
with extended-group ideas. The $\mu$-dependence of the molecular parameters that are used in this model
are derived and the most important ones of these are validated using the spectroscopic data of
different isotopologues of methylamine. We find a significant enhancement of the sensitivity 
coefficients due to energy cancellations between internal rotational, overall rotational and 
inversion energy splittings. The sensitivity coefficients of the different transitions range 
from $-19$ to $+24$. The sensitivity coefficients of the 78.135, 79.008, and 89.956\,GHz transitions 
that were recently observed in the disk of a $z=0.89$ spiral galaxy located in front of the quasar 
PKS 1830-211 [S.~Muller~\emph{et al.} Astron. Astrophys. {\bf 535}, A103 (2011)]  were calculated to be $-0.87$ 
for the first two and $-1.4$ for the third transition, respectively.
From these transitions a preliminary upper limit for a variation 
of the proton to electron mass ratio of  $\Delta \mu/\mu< 9\times10^{-6}$ is deduced.
\end{abstract}

\pacs{06.20.Jr, 33.15.-e, 98.80.-k}

\maketitle

\section{Introduction\label{sec:introduction}} 

Recently, it was shown that transitions between accidently degenerate levels that 
correspond to different motional states in polyatomic molecules are very sensitive
to  a possible variation of the proton-to-electron mass ratio, $\mu=m_{p}/m_{e}$.  
Kozlov \emph{et al.}~\cite{Kozlov2011PRA} showed that transitions that convert rotational
motion into inversion motion, and vice versa, in the different isotopologues 
of hydronium (\ce{H_{3}O+}) have $K_{\mu}$ coefficients ranging from $-219$ to $+11$
\footnote{In Kozlov~\emph{et al.}~\cite{Kozlov2011PRA}, $\mu$ is defined as the electron-to-proton
mass ratio, consequently, the sensitivity coefficient used in that work is 
minus times the sensitivity coefficient used here, $Q_{\mu}$=$-K_{\mu}$.}. 
Similarly, Jansen \emph{et al.}~\cite{Jansen2011PRL,Jansen2011PRA} 
and Levshakov~\emph{et al.}~\cite{Levshakov2011} showed that transitions that 
convert internal rotation into overall rotation in the different isotopologues of methanol have 
$K_{\mu}$ coefficients ranging from $-88$ to $+330$. Here, the sensitivy coefficient, $K_{\mu}$, is defined by
\begin{equation}
\frac{\Delta \nu}{\nu}=K_{\mu}\frac{\Delta \mu}{\mu}.
\label{eq:Kmu}
\end{equation} 

\noindent
For comparison, pure rotational transitions have $K_{\mu}=-1$, while pure vibrational
transitions have $K_{\mu}=-\tfrac{1}{2}$ and pure electronic transitions have $K_{\mu}=0$.

\begin{figure}[tb]
\includegraphics[width=\columnwidth]{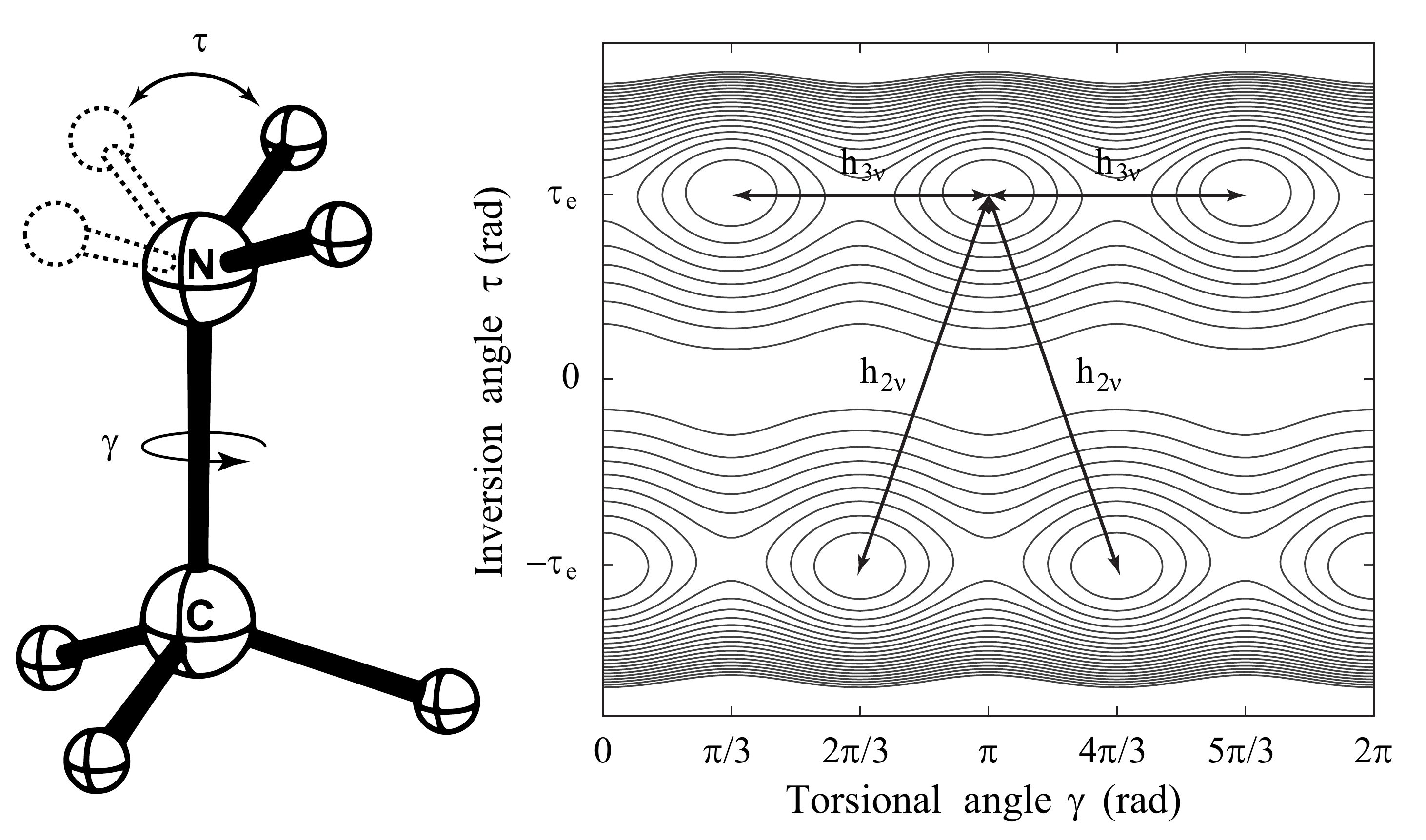}
\caption{Schematic representation of methylamine and variation of the potential
energy of methylamine as function of the relative rotation, $\gamma$, of the
\ce{CH3}-group with respect to the amine group about the \ce{CN} bond and the angle,
$\tau$, of the two hydrogen atoms of the \ce{NH2}-group with respect to the \ce{CN}
bond. The two large amplitude motions, corresponding to inversion ($h_{2v}$)
and hindered rotation ($h_{3v}$) are schematically indicated by the arrows.
Note that inversion of the \ce{NH2}-group is accompanied by a $\pi/3$ rotation
about the \ce{CN} bond of the \ce{CH3}-group with respect to the amine group.
\label{fig:potential_and_structure}}
\end{figure}

Accidental degeneracies between different motional states in polyatomic molecules are 
likely to occur if the energies associated with the different types of motions 
are similar. In this paper, we present a calculation of the sensitivity 
coefficients for microwave transitions in methylamine (\ce{CH_{3}NH_{2}}). 
Methylamine is an interesting molecule for several 
reasons: (i) it displays two large amplitude motions; hindered internal rotation of the methyl (\ce{CH3})
group with respect to the amino group (\ce{NH2}), and tunneling associated with
wagging of the amino group. 
The coupling between the internal rotation and overall rotation in methylamine is rather strong
resulting in a strong dependence of the torsional energies on the $K$ quantum number, which
is favorable for obtaining large enhancements of the $K_{\mu}$-coefficients~\cite{Jansen2011PRA}. 
(ii) Methylamine is a relatively small and stable molecule that is abundantly present in our galaxy 
and easy to work with in the laboratory. Recently it was also detected in the disk of a 
high redshift ($z=0.89$) spiral galaxy located in front of the quasar PKS 1830-211\cite{Muller2011}. 

This paper is organized as follows. In Section~\ref{Sec:Hamiltonian}, we introduce the 
effective Hamiltonian used for calculating the level energies in the vibrational 
ground state of methylamine. In Section~\ref{Sec:Scaling} we derive how the constants
that appear in this Hamiltonian scale with $\mu$. Finally, in Section~\ref{Sec:Results}
we use the Hamiltonian and the scaling relations to determine the sensitivity coefficients
of selected transitions.  

\section{Hamiltonian and energy level structure}
\label{Sec:Hamiltonian}

Methylamine, schematically depicted in Fig.~\ref{fig:potential_and_structure}, is a 
representative of molecules exhibiting two coupled large-amplitude motions, the torsional motion 
of a methyl group and the wagging (or inversion) motion of an amine group. 
A combination of intermediate heights of the potential barriers with a leading role 
of the light hydrogen atoms in the large-amplitude motions results in relatively large tunneling 
splittings even in the ground vibrational state. On the right-hand side of 
Fig.~\ref{fig:potential_and_structure}, a contour plot of the potential energy is shown with the relative 
angle between the methyl and the amino group, $\gamma$, on the horizontal axis and the angle between the \ce{NH_2}-plane 
and the \ce{CN}-bond, $\tau$, on the vertical axis. The methyl torsion motion is indicated with the arrow 
labeled by $h_{3v}$ whereas the amino wagging motion is indicated with the arrow labeled by $h_{2v}$. 
From the contour plot, it is seen that amino wagging motion of the \ce{NH_2}-group is accompanied by 
a $\pi/3$ rotation of the \ce{CH_3}-group about the \ce{CN} bond with respect to the \ce{NH_2} group. 
Consequently, the amino wagging motion is strongly coupled to the hindered methyl top internal rotation 
resulting in a rather complicated computational problem.

\begin{figure*}[tb]
\includegraphics[width=\textwidth]{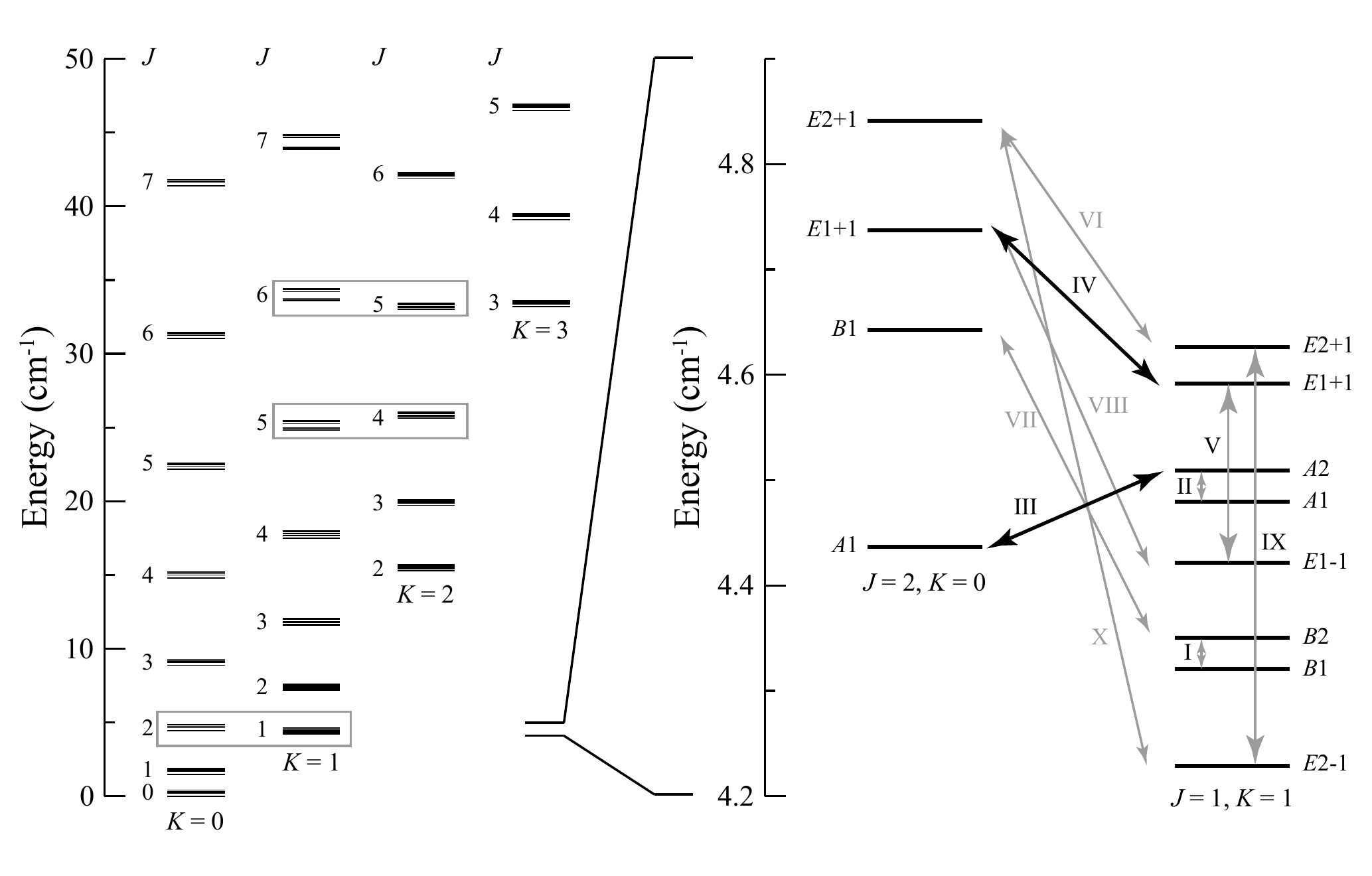}
\caption{Energy of the lowest rotational levels in the ground state of methylamine (\ce{^{12}CH_{3}^{14}NH_{2}}). 
The levels are denoted by $J$, $K$ and the overall symmetry.
The panel on the right hand side of the figure shows an enlarged view of the 
$J=2,K=0$ and $J=1,K=1$ levels, with all symmetry allowed transitions assigned with roman numerals.
The sensitivity of these transitions are listed in Table~\ref{Tab:ColdTrans}. The two transitions
that are designated with bold arrows and are labeled by {\sc iii} and {\sc iv} 
have sensitivity coefficients equal to $K_{\mu}=-19$ and $K_{\mu}=2$, respectively.
\label{Fig:levelscheme}}
\end{figure*}

In Fig.~\ref{Fig:levelscheme} the lowest rotational levels of the ground vibrational state of
\ce{CH3NH2} are shown. The level ordering resembles that of a near-prolate
asymmetric top molecule. In addition to the usual asymmetric splitting, every
$J$, $K$ level is split due to the different tunneling motions. The internal
rotation tunneling splits each rotational level into one doubly degenerate
and one nondegenerate sublevel. Each of these sublevels are further split
into two due to the inversion motion. Together, this results in eight
levels with overall symmetry $A1$, $A2$, $B1$, $B2$, $E1+1$, $E2+1$, $E1-1$ and $E2-1$ 
for $K>0$ and four levels for $K=0$. The $+1$ and $-1$ levels in the $E1$ and $E2$
symmetry species, correspond to $K>0$ and $K<0$ respectively. Because
of nuclear-spin statistics, in the ground vibrational state the
nondegenerate levels of $J=$even, $K=0$ are only allowed to possess the
overall symmetry $A1$, $B1$, whereas levels with $J=$odd, $K=0$ are only allowed
to possess the overall symmetry $A2$, $B2$. The $K=0$ doubly degenerate levels of
$E1$ and $E2$ symmetry are denoted by $+1$ levels, i.e. by $E1+1$, $E2+1$
levels. The exact ordering of the different symmetry levels within a
certain $J$, $K$ level is determined by the relative contributions of the $h_{3v}$ and
$h_{2v}$ parameters (see for example Fig.~3 of Ref.~\cite{Ilyushin_JCP2010}). 
The internal motions are strongly coupled to the overall rotation 
resulting in a strong dependence of the torsional-wagging energies on the $K$ quantum number. 
Thus the level ordering may differ from one $K$-ladder to another. 
This turns out to be important for obtaining large enhancement factors, as it may result 
in closely spaced energy levels with a different functional dependence on $\mu$ which
are connected by a symmetry allowed transition.

The panel on the right hand side of Fig.~\ref{Fig:levelscheme} shows an enlarged view of the 
$J=2$, $K=0$ and $J=1$, $K=1$ levels, with all symmetry allowed transitions assigned with roman numerals. 
Note that transitions with $\Delta J=0$ in the $K=0$ manifold are not allowed.
The transitions labeled by {\sc iii,iv,vi,vii,viii} and {\sc x} are of particular interest
as these connect the closely spaced levels of different $K$-manifolds and have
an enhanced sensitivity to a variation of $\mu$. A similar enhancement occurs for transitions 
between the $J=5$, $K=1$, and $J=4$, $K=2$ levels as well as for transitions between the $J=6$, $K=1$, 
and $J=5$, $K=2$ levels. In what follows, we will outline the procedure to calculate the sensitivities
of these transitions. The resulting sensitivity coefficients are presented in Table~\ref{Tab:Lovas}
and Table~\ref{Tab:ColdTrans} and discussed in Sec.~\ref{Sec:Results}.

In the present work, we use the group-theoretical high-barrier tunneling formalism developed for 
methylamine by Ohashi and Hougen~\cite{Ohashi&Hougen1987}, which is capable of reproducing observations 
of the rotational spectrum of the ground vibrational state of \ce{CH_{3}NH_{2}} to within a few tens 
of kilohertz~\cite{Ilyushin2005,Ilyushin2007}. The high-barrier formalism assumes that the molecule is 
confined to one of $n$ equivalent equilibrium potential minima for many vibrations, but that it 
occasionally tunnels from one of these $n$ minima to another. The formalism fits in between the 
infinite-barrier approximation, where no tunneling splittings are observed, and the low-barrier 
approximation, where the present formalism breaks down. A backward rotation of the whole molecule is 
introduced to cancel the angular momentum generated by one of the large amplitude motions --
the so-called internal axis method -- requiring the usage of extended group ideas. The reader is 
referred to Refs.~\cite{Ohashi&Hougen1987, Ilyushin2005, Ilyushin2007, Hougen&DeKoven1983, 
Ilyushin_JCP2010} for a detailed description of the high-barrier tunneling formalism and the used Hamiltonian.

Table~\ref{Tab:MolecularConstants} lists the molecular constants used 
in our calculations. It includes three types of parameters: `non-tunneling' or pure 
rotational parameters; parameters associated with pure methyl torsion motion (odd numerical subscripts $n$); 
and parameters associated with the \ce{NH2} wagging motion (even numerical subscripts $n$).
The obtained $\mu$-scaling relations for the different parameters of the high-barrier tunneling formalism of methylamine 
are listed in the rightmost column of Table~\ref{Tab:MolecularConstants}. In the next sections, we will discuss the 
scaling relations for the lowest order parameters, the scaling relations for the 
higher order parameters, and the problems encountered in determining these, are discussed in the 
supplementary online material~\cite{onlinematerial}.

\begingroup
\squeezetable
\renewcommand{\arraystretch}{1.1}
\begin{table*}%[H] add [H] placement to break table across pages
\caption{Molecular parameters, $P_{s}$, of the ground torsional state of methylamine \ce{CH3NH2}~\cite{Ilyushin2005}, 
and their sensitivity to a variation of the proton-to-electron mass ratio $\mu$ defined as  
$K^{P_{s}}_{\mu} = \frac{\mu}{P_{s}}\frac{\partial P_{s}}{\partial \mu}$. All molecular parameters are in MHz, except $\rho$ and $\rho_{K}$, 
which are dimensionless. \label{Tab:MolecularConstants}}
\begin{ruledtabular}
\begin{tabular}{>{$}l<{$} D{.}{.}{3} D{.}{.}{15} >{$}l<{$} D{.}{.}{3} D{.}{.}{15} >{$}l<{$} D{.}{.}{3} D{.}{.}{15}}
\multicolumn{3}{c}{Rotation\footnote[1]{These parameters do not involve tunneling motions.}}&\multicolumn{3}{c}
{Inversion\footnote[2]{These parameters arise from the \ce{NH2} inversion tunneling motion.}} &\multicolumn{3}{c}
{Torsion\footnote[3]{These parameters arise from the \ce{CH3} torsional tunneling motions.}} \\\cline{1-3}\cline{4-6}\cline{7-9}
\\[-1.9ex]
 & \multicolumn{1}{c}{$K^{P_{s}}_{\mu}$} &  &  & \multicolumn{1}{c}{$K^{P_{s}}_{\mu}$} &  &  & \multicolumn{1}{c}{$K^{P_{s}}_{\mu}$} &\\
\bar{B}     &-1& 22\,169.36636(30)& h_{2v}    &-5.5& -1\,549.18621(77) & h_{3v}         &-4.7& -2\,493.5140(12)      \\
A-\bar{B}   &-1& 80\,986.3823(11) & h_{4v}    &-8.2& 2.73186(96)       & h_{5v}         &-8.8& 2.88398(55)           \\
B-C         &-1& 877.87717(53)    & h_{2J}    &-5.5& 0.101759(11)      & h_{3J}         &-4.7& -0.052546(20)         \\
D_J	    &-2& 0.0394510(18)    & h_{2K}    &-5.5& 1.73955(16)       & h_{5J}         &-8.8& 0.0002282(55)         \\
D_{JK}	    &-2& 0.170986(15)     & h_{4K}    &-8.2& -0.004778(37)     & h_{3K}         &-4.7& 1.16676(22)           \\
D_K	    &-2& 0.701044(24)     & h_{2JJ}   &-6.5& -0.000005466(88)  & h_{5K}         &-8.8& -0.002667(73)         \\
\delta_J    &-2& 0.00175673(17)   & h_{2KK}   &-6.5& -0.0009016(63)    & h_{3JJ}        &-5.7& -0.000017296(44)      \\
\delta_K    &-2& -0.33772(13)     & h_{2JK}   &-6.5& -0.00015400(94)   & h_{3KK}        &-5.7& -0.0002995(42)        \\
\Phi_{J}    &-3& -0.0000000485(16)& h_{2JKK}  &-7.5& 0.0000001923(56)  & h_{3JJK}       &-6.7& -0.00000004702(67)    \\
\Phi_{JK}   &-3& 0.000002442(50)  & q_2	      &-5.5& 21.54923(52)      & f_3            &-4.7& -0.173439(24)         \\
\Phi_{KJ}   &-3& -0.00000855(10)  & q_4	      &-8.2& -0.03071(20)      & f_{3J}         &-5.7& -0.00000261(13)       \\
\Phi_K	    &-3& 0.00003322(29)   & q_{2J}    &-6.5& -0.0037368(45)    & f_{3K}         &-5.7& -0.0001359(32)        \\
\phi_K	    &-3& 0.0002366(48)    & q_{2K}    &-6.5& -0.019676(43)     & f_{3JK}        &-6.7& -0.0000000646(27)     \\
	    &  &                  & q_{2JJ}   &-7.5& 0.000002098(62)   & f^ {(2)}_3     &-5.7& -0.000003021(89)      \\
	    &  &	          & q_{2KK}   &-7.5& 0.00001023(54)    & f^ {(2)}_{3J}  &-6.7& 0.00000000220(13)     \\
\rho	    & 0& 0.64976023(13)   & f_2	      &-5.5& -0.096739(38)     &                &    &	                     \\
\rho_K      &-1& -0.0000011601(77)& f_4	      &-8.2& 0.0002153(39)     &                &    &	                     \\
	    &  &                  & f_{2J}    &-6.5& 0.000004452(67)   &                &    &	                     \\
	    &  &                  & f_{2K}    &-6.5& 0.001188(37)      &                &    &	                     \\
	    &  &                  & f_{2KK}   &-7.5& -0.000001600(47)  &                &    &	                     \\
	    &  &                  & f^{(2)}_2 &-6.5& -0.000002443(55)  &                &    &	                     \\
	    &  &                  & r_2	      &-5.5& 10.979(37)        &                &    &	                     \\
	    &  &                  & r_{2K}    &-6.5& -0.7206(73)       &                &    &	                     \\

\end{tabular}
\end{ruledtabular}
\end{table*}
\renewcommand{\arraystretch}{1.1}
\endgroup

\section{Scaling relations of the molecular parameters}
\label{Sec:Scaling}

We will use two different approaches for determining the $\mu$-dependence of the molecular 
constants that appear in the Hamiltonian: 

(i) The first approach is based on the fact that the tunneling model essentially assumes 
that for each large-amplitude tunneling motion the system-point travels along some path in coordinate 
space. In zeroth approximation, we may represent each large amplitude motion as a one-dimensional 
mathematical problem after parameterizing the potential along the path and the effective mass that moves 
along it. Thus, for each large amplitude motion, we will set up a Hamiltonian that contains one 
position coordinate and its momentum conjugate. The parameters of this one dimensional 
Hamiltonian may be connected with the observed splittings which are fitting parameters of the high-barrier 
tunneling formalism. The parameters of the one-dimensional Hamiltonians are functions of the moments of 
inertia and the potential barrier only, and their $\mu$-dependence can be found in a similar fashion as was done 
for methanol and other internal rotors~\cite{Jansen2011PRL,Jansen2011PRA}. Application of this approach is 
straightforward in the case of the leading tunneling parameters of methylamine but 
some ambiguities appear for the $J$ and $K$ dependences of the main terms, because
there are several ways of representing these dependences in a one dimensional model.

(ii) In the second approach, we use the spectroscopic data of different isotopologues 
of methylamine to estimate the dependence of the tunneling constants. 
In analogy with methanol, we expect the tunneling splittings to follow the formula~\cite{Jansen2011PRL}:

\begin{equation}
W_{\mathrm{\it{splitting}}} = \frac{a_0}{\sqrt{I_{\mathrm{\it{red}}}}} e^{-a_1 \sqrt{I_{\mathrm{\it{red}}}}}.
\label{eq:WKB}
\end{equation}

\noindent  
This formula originates from the semi-classical (Wentzel-Kramers-Brillouin (WKB)) approximation 
that assumes that the effective tunneling mass, represented by $I_{\mathrm{\it red}}$, changes
with isotopic substitution, but that the barrier between different wells remains unchanged. 
This expression was successfully applied to the $J=0$, $K=0$ $A-E$ 
splittings and the $J=1$, $\vert K\vert=1$ splittings in methanol~\cite{Jansen2011PRL}. 
In methylamine, the $h_{nv}$ parameters correspond to the splittings in the $J=0$, $K=0$ due to 
tunneling between framework $\vert 1\rangle$ and framework $\vert n\rangle$ (the
set of frameworks represent the equivalent potential wells between which
the system can tunnel), and application of the WKB approach to these parameters is straightforward. 
Moreover, since in fact all tunneling parameters in methylamine 
may be related to the same type of overlap integral as the $h_{nv}$ parameters, we may expect that the isotopologue 
dependence of all tunneling terms
can be described by Eq.~(\refeq{eq:WKB}). Unfortunately, ambiguities appear again when we apply this approach
to higher order terms in the methylamine Hamiltonian. These ambiguities are connected to the fact that vibrational 
basis set functions $\vert n \rangle$ localized near various minima are not orthogonal, but in fact have nonzero 
overlap integrals with each other. The correlation problems 
that arise in the high-barrier tunneling formalism due to nonorthogonality of the basis functions are discussed in 
some detail in Ref.~\cite{Ohashi&Hougen1985}. The main consequence which affects the isotopologue approach is 
that there may be `leakage' from one parameter to another; each fitted parameter appears as a sum of the 
`true' parameter value plus a small linear combination of all other parameters with a coefficient that goes to 
zero when the overlap integral goes to zero. While this effect should be insignificant for the main tunneling 
parameters of methylamine, it may be important for higher order terms because even a small `leakage' of the 
low order parameters may be comparable in magnitude with the `true' values of the higher order parameter. 

In order to verify the mass dependence coefficients for the parameters of the methylamine Hamiltonian, we have 
refitted available data on the \ce{CH3ND2}~\cite{Takagi1971}, \ce{CD3NH2}~\cite{Kreglewski1990CD3NH2} and
\ce{CD3ND2}\cite{Kreglewski1990CD3ND2} isotopologues of methylamine using the high-barrier tunneling formalism. 
Unfortunately, the amount of data available in the literature was rather limited; 66 transitions for 
\ce{CH3ND2}~\cite{Takagi1971}, 41 transition for \ce{CD3NH2}~\cite{Kreglewski1990CD3NH2} and 49 transitions 
for \ce{CD3ND2}\cite{Kreglewski1990CD3ND2}. Therefore, many of the higher order terms were not determined in the 
fits, while some low order parameters were determined with a few significant digits only. As a result, it was
possible to obtain the $\mu$-dependence of the main tunneling parameters $h_{2v}$ and $h_{3v}$ only.
In order to obtain information on higher order terms,
we have undertaken a new investigation of the \ce{CH3ND2} spectrum with the Kharkov millimeter wave spectrometer. 
The newly obtained dataset for \ce{CH3ND2} 
contains 614 transitions, comparable to the number of microwave transitions available for \ce{CH3NH2} 
(696 transitions). The \ce{CH3NH2} and \ce{CH3ND2} fits have an almost equal number of varied parameters 
and obtained similar weighted root-mean-square deviations. The results of the \ce{CH3ND2} investigation
will be published elsewhere~\cite{Ilyushin:inpreparation}, here we will use only those results necessary
for obtaining the scaling relations.

\subsection{Pure Rotational Constants} 

The pure rotational or `nontunneling' parameters in the model are connected to the usual moments of inertia of the 
molecule and to the centrifugal distortion parameters. Therefore, we will assume the same $\mu$-dependence 
for these parameters as used for methanol~\cite{Jansen2011PRA}. 

\subsection{\ce{CH3} torsion and the $h_{3v}$ parameter} 

The $h_{3v}$ parameter in the high barrier-tunneling formalism corresponds to a pure torsion motion. 
The quantity $\vert 3h_{3v}\vert$ may be related to the usual $E$-$A$ internal rotation splitting in a 
molecule that contains a group of $C_{3v}$ symmetry. Assuming that the potential barrier is described
by a cosine function and taking the moment of inertia of the methyl top to represent the mass that  
tunnels, we may set up a one-dimension internal rotation Hamiltonian

\begin{equation}
H_{\mathrm{\it{tors}}} = F_{\gamma}p_{\gamma}^{2} + \frac{V_{n}}{2} \left(1 - \cos{n\gamma}\right),
\label{Eq:Hamiltonian-1d}
\end{equation}

\noindent
with $n=3$ for a threefold barrier, $p_{\gamma}= -i\partial/\partial\gamma$ is the angular momentum operator 
associated to the internal rotation coordinate, $F_{\gamma}$ is the internal rotation parameter and $V_{3}$ the 
barrier height. Using a value for $F_{\gamma}$ derived from the molecular constants, we may fit the barrier 
height $V_{3}$ to the observed value for $\vert 3h_{3v}\vert$ and estimate the 
$\mu$-dependence of $h_{3v}$. 

In the used axis system, the off-diagonal contribution to the inertia tensor is represented by 
the $s_{1}$ parameter. For methylamine, this parameter is set to zero as it is not required by the fit. 
Thus, we may assume that the methyl top axis coincides with the principal axis $a$, and $\rho=I_{\gamma}/I_{a}$, and 
$F_{\gamma}=C_{\mathrm{\it{conv}}}/((1 -\rho)I_{\gamma})$, with $C_{\mathrm{\it{conv}}}$ being a conversion factor 
($C_{\mathrm{\it{conv}}}= 16.8576291\text{\,amu\,\AA\,cm$^{-1}$}$). Using values for $\rho$ and $I_{a}$
(recalculated from rotational parameters) from Table~\ref{Tab:MolecularConstants}, we obtain 
$I_{\gamma} = 3.18\text{\,amu\,\AA$^2$}$ and $F_{\gamma} = 15.12\text{\,cm$^{-1}$}$ (\emph{ab initio} value 
$15.1684\text{\,cm$^{-1}$}$~\cite{Smeyers1996}). The value for $I_{\gamma}$ is close to the expected one 
which supports the validity of the present analysis. Now, using this value for $F_{\gamma}$ and the value for $h_{3v}$ from 
Table~\ref{Tab:MolecularConstants}, a fit to Eq.~(\refeq{Eq:Hamiltonian-1d}) yields the effective barrier height
$V_{3}=683.7$\,cm$^{-1}$ (\emph{ab initio} value 708.64\,cm$^{-1}$~\cite{Smeyers1996}). 
The one-dimensional model with this value for $V_{3}$ predicts values for the first torsional band and 
the $A-E$ splitting in the first excited torsional state that are in a good agreement with the observed values (269\,cm$^{-1}$ versus 264\,cm$^{-1}$~\cite{Kreglewski1992} 
for the band origin and 186\,GHz versus 180\,GHz~\cite{Kreglewski1992} for the splitting in the $\nu_{t}$=1). 
All this indicates that the one dimensional model is physically sound and sufficiently accurate for our purposes.

Finally, we obtain the $\mu$-dependence of $h_{3v}$ via 

\begin{align}
K_{\mu}^{h_{3v}} &= \frac{\mu}{h_{3v}} \frac{\partial(h_{3v})}{\partial \mu}\nonumber\\
&= -\frac{F_{\gamma}}{h_{3v}} \frac{\partial(h_{3v})}{\partial F_{\gamma}},
\label{eq:Kh3v}
\end{align}

\noindent
where we have used the fact that $F_{\gamma}$ scales as $\mu^{-1}$, i.e., we assume that the neutron mass 
has a similar variation as the proton mass. The numerical evaluation $\partial (h_{3v})/\partial F_{\gamma}$ 
using Eq.~(\refeq{eq:Kh3v}) yields $K_{\mu}^{h_{3v}}=-4.66$.

\begin{figure}[tb]
\includegraphics[width=0.9\columnwidth]{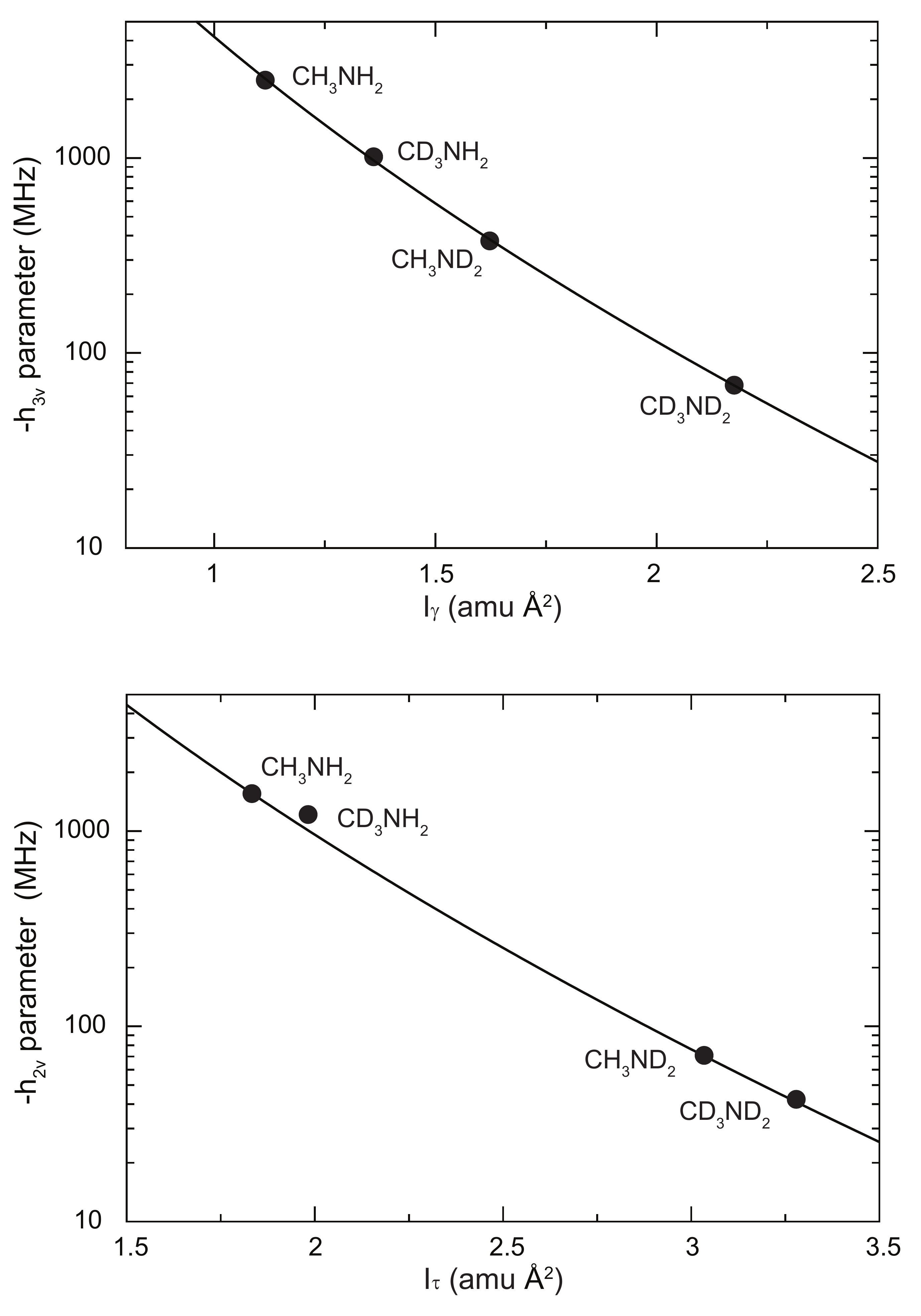}
\caption{$h_{2v}$ and $h_{3v}$ parameters as a function of the reduced moment of inertia 
for the torsional and inversion motion for four different isotopologues 
of methylamine. The solid lines are fits according to Eq.~{\eqref{eq:WKB}}
through the values of \ce{CH3NH2} and \ce{CH3ND2}.  
\label{fig:isotopologues}}
\end{figure}

In the upper panel of Fig.~\ref{fig:isotopologues}, the value of the $h_{3v}$ parameter is plotted as a function 
of the reduced moment of inertia, $I_{\mathrm{\it{red}}}^{\gamma} = C_{\textit{conv}}/F_{\gamma}$ for 4 
different isotopologues of methylamine. As mentioned, the quantity $\vert 3h_{3v}\vert$ corresponds to 
the usual $A-E$ internal rotation splitting in a methyl top molecule, hence, we expect the tunneling splitting 
to follow Eq.~(\refeq{eq:WKB}). The solid line in the upper panel of Fig.~\ref{fig:isotopologues}, corresponds to 
$a_{0} = {10.3}~{\text{THz}}~{(\text{amu \AA$^2$)}^{1/2}}$ and $a_{1}={7.84}~{\text{(amu \AA$^2$)}^{-1/2}}$, 
obtained using the \ce{CH3NH2} and \ce{CH3ND2} data. The reduced moment of inertia is directly 
proportional to $\mu$. Thus, the sensitivity coefficient is given by: 
 
\begin{align}
K_{\mu}^{h_{3v}} &= 
\frac{I_{\mathrm{\it{red}}}^{\gamma}}{h_{3v}} \frac{\partial(h_{3v})}{\partial I_{\mathrm{\it{red}}}^{\gamma}}\nonumber\\
&= - \frac{1}{2}-\frac{a_{1}\sqrt{I_{\mathrm{\it{red}}}^{\gamma}}}{2}.
\label{eq:Kh3v_isotopolgues}
\end{align}
 
\noindent
From the above expression, we find for the $h_{3v}$ parameter of \ce{CH3NH2} a sensitivity coefficient of 
$K_{\mu}^{h_{3v}}=-4.64$, in excellent agreement with the value found from the one-dimensional Hamiltonian model. 

\subsection{Inversion and the $h_{2v}$ parameter} 

The interpretation of the $h_{2v}$ parameter in terms of an effective mass moving in a 
one-dimensional effective potential is not straightforward. For instance, \emph{ab initio} calculations of 
the kinetic parameter for the inversion motion in the equilibrium geometry range 
from 9.6017\,cm$^{-1}$~\cite{Smeyers1996} to 26.7291\,cm$^{-1}$~\cite{Smeyers1998}, 
while the barrier height in different studies varies from 1686\,cm$^{-1}$\cite{Tsuboi1966} 
to 2081\,cm$^{-1}$\cite{Sztraka1987}. Since the system needs to tunnel six times in order to return to its initial 
configuration, we will treat this large amplitude motion as a six-fold periodic well 
problem, following Ohashi~\emph{et al.}~\cite{Ohashi1988}. Furthermore, we assume that the potential along the path can be 
represented by a rapidly converging Fourier series. Thus, we use Eq.~(\refeq{Eq:Hamiltonian-1d}) with $\gamma$
replaced by $\tau$ and $n=6$ as a zeroth order model. 
The effective inversion-torsion constant $F_{\tau}$ and barrier height $V_{6}$ can be determined from the splittings in 
the ground state and \ce{NH2} wagging band origin (780\,cm$^{-1}$~\cite{Tsuboi1964}). 
From this, we obtain $F_{\tau}$=9.19\,cm$^{-1}$ and $V_{6}$=2322\,cm$^{-1}$, close to the values obtained by 
Ohashi~\emph{et al.}~\cite{Ohashi1988}. Following the same procedure as for $h_{3v}$, 
we obtain the $\mu$  dependence of $h_{2v}$, $K_{\mu}^{h_{2v}}=-5.49$.

In the lower panel of Fig.~\ref{fig:isotopologues}, the value of the $h_{2v}$ parameter is plotted 
as a function of the reduced moment of inertia, $I_{\mathrm{\it{red}}}^{\tau} = C_{\textit{conv}}/F_{\tau}$ 
for 4 different isotopologues of methylamine. The solid line in Fig.~\ref{fig:isotopologues} corresponds to 
$a_{0} = {44.4}~{\text{THz}}~{(\text{amu \AA$^2$)}^{1/2}}$ and 
$a_{1} = {7.35}~{\text{(amu \AA$^2$)}^{-1/2}}$ obtained using the \ce{CH3NH2} and \ce{CH3ND2} data.
From this fit, we find for the $h_{2v}$ parameter of \ce{CH_{3}NH_{2}} a sensitivity coefficient equal to 
$K_{\mu}^{h_{2v}}=-5.48$, again in excellent agreement with the one-dimensional 
Hamiltonian model.

\subsection{$q_{2}$ and $r_{2}$ parameters} 

The linear terms $q$ and $r$ correspond to the interaction of components of the total angular momentum 
with the angular momentum generated in the molecule-fixed axis system by the two large amplitude motions. 
In methylamine, $q_{2}$ and $r_{2}$ represent the interaction of the angular momentum 
generated by the \ce{NH2} inversion and the `corrective' $\pi$/3 rotation of the \ce{CH3} group with the 
$J_{z}$ and $J_{y}$ components of the total angular momentum, respectively. It can be shown in different ways that $q_{2}$ has the same 
dependence on $\mu$ as $h_{2v}$. For instance, it follows from a study of the correlations 
between the $q_{2}$, $q_{3}$ and $\rho$ parameters carried out by Ohashi and Hougen~\cite{Ohashi&Hougen1987}.
In methylamine, two possible choices exist for $\rho$. $\rho$ can be chosen such that Coriolis coupling due to 
the inversion plus corrective rotation is eliminated ($q_{2}$ fixed to zero), or such that Coriolis coupling 
due to the internal rotation of the \ce{CH3} group is eliminated ($q_{3}$ fixed to zero). These two choices 
result in a difference $\Delta\rho  =(3/\pi)q_{2}/h_{2v}$~\cite{Ohashi&Hougen1987}. Since $\rho$ is in both cases 
a (dimensionless) ratio between different moments of inertia and independent of $\mu$, the above equation 
implies that $q_{2}$ and $h_{2v}$ should have the same $\mu$-dependence. 

From the \ce{CH3ND2} isotopologue 
data, a sensitivity coefficient $K_{\mu}^{q_{2}}=-5.53$ was found, in good agreement with the
$K_{\mu}^{q_{2}}=-5.50$ obtained from the one-dimensional model and close to the value for $K_{\mu}^{h_{2v}}$.
The $r_{2}$ term is expected to have the same $\mu$-dependence as $q_{2}$.
We were not able to check the isotopologue dependence for this term, since it was not required by 
the \ce{CH3ND2} fit.

\subsection{Higher order terms}

The $\mu$-dependence of the higher order terms, including the $J$ and $K$ dependences 
of the $h_{2v}$ and $h_{3v}$ parameters, was determined in a similar fashion (see the online material to this paper
~\cite{onlinematerial}). Unfortunately, some ambiguities and discrepancies between the different 
approaches appeared in the determination of the scaling relations for some higher order terms, 
which is reflected by the rather large error for these parameters (see Sec.~\ref{Sec:Results}). 
This is not a serious concern as the higher order tunneling parameters only marginally affect the 
$K_{\mu}$ coefficients of the considered transitions.

\begingroup
\squeezetable
\renewcommand{\arraystretch}{1.1}
\begin{table*}
\caption{Transitions in methylamine (\ce{CH3NH2}) that are detected in 
astrophysical objects in our local galaxy as listed in Lovas \emph{et al.}~\cite{Lovas2004}.
The fourth column lists the transition strength multiplied by 
the electric dipole moment, $\mu_{e}$, squared.
The last column lists the sensitivity of the transitions
to a possible variation of the proton-to-electron mass ratio. 
\label{Tab:Lovas}}
\begin{tabular}{ll}
\begin{tabular}[t]{ >{$}c<{$}>{$}c<{$}>{$}l<{$} c
 >{$}c<{$}>{$}c<{$}>{$}l<{$} D..{6} D..{6} D..{6}}
\hline\hline
\multicolumn{3}{c}{Upper state} & &
\multicolumn{3}{c}{Lower state} &
\multicolumn{1}{c}{Transition (MHz)} &
\multicolumn{1}{c}{S$\mu_{e}^{2}$ (D$^2$)} &
\multicolumn{1}{c}{$K_{\mu}$}\bigstrut\\\cline{1-3}\cline{5-7}
\\[-1.9ex]
J & K & \text{Sym} & & J & K & \text{Sym} & & &  \\
\hline
\\[-1.8ex]
    2 &    0 &  B1 & &    1 &    1 &  B2 & 8\,777.827 & 0.779 & -2.14(6)
                    \\
    5 &    1 &  B1 & &    5 &    0 &  B2 & 73\,044.474 & 9.024 & -0.86(3)
                    \\
    4 &    1 &  B2 & &    4 &    0 &  B1 & 75\,134.858 & 7.290 & -0.87(3)
                    \\
    3 &    1 &  B1 & &    3 &    0 &  B2 & 76\,838.932 & 5.611 & -0.87(3)
                    \\
    1 &    1 &  B1 & &    1 &    0 &  B2 & 79\,008.693 & 2.373 & -0.87(3)
                    \\
    5 &    1 &  A1 & &    5 &    0 &  A2 & 83\,978.941 & 9.024 & -1.47(4)
                    \\
    2 &    1 &  E1+1 & &    2 &    0 &  E1+1 & 84\,598.202 & 1.065 & -1.14(3)
                    \\
    4 &    1 &  A2 & &    4 &    0 &  A1 & 86\,074.729 & 7.290 & -1.45(4)
                    \\
    3 &    1 &  A1 & &    3 &    0 &  A2 & 87\,782.494 & 5.613 & -1.45(4)
                    \\
    2 &    0 &  B1 & &    1 &    0 &  B2 & 88\,667.906 & 0.189 & -1.00(3)
                    \\
    2 &    0 &  E2+1 & &    1 &    0 &  E2+1 & 88\,668.681 & 0.189 & -1.00(3)
                        \\
    2 &    0 &  E1+1 & &    1 &    0 &  E1+1 & 88\,669.543 & 0.188 & -1.00(3)
                        \\
    2 &    0 &  A1 & &    1 &    0 &  A2 & 88\,669.626 & 0.188 & -1.00(3)
                        \\
    8 &    2 &  E1-1 & &    8 &    1 &  E1+1 & 219\,151.221 & 3.519 &
-0.84(3)
                    \\
    7 &    0 &  B2 & &    6 &    1 &  B1 & 220\,826.705 & 4.295 & -1.05(3)
                    \\
    9 &    2 &  E2+1 & &    9 &    1 &  E2+1 & 220\,888.443 & 7.496 &
-0.94(3)
                    \\
    5 &    0 &  E2+1 & &    4 &    0 &  E2+1 & 221\,527.438 & 0.472 &
-1.00(3)
                        \\
    5 &    0 &  E1+1 & &    4 &    0 &  E1+1 & 221\,530.404 & 0.470 &
-1.00(3)
                        \\
    5 &    0 &  B2 & &    4 &    0 &  B1 & 221\,530.481 & 0.473 & -1.00(3)
                        \\
    5 &    0 &  A2 & &    4 &    0 &  A1 & 221\,536.285 & 0.470 & -1.00(3)
                        \\
    5 &    2 &  E2+1 & &    4 &    2 &  E2+1 & 221\,717.567 & 0.395 &
-1.00(3)
                        \\
    5 &    2 &  E1+1 & &    4 &    2 &  E1+1 & 221\,721.771 & 0.396 &
-1.00(3)
                        \\
    5 &    2 &  E1-1 & &    4 &    2 &  E1-1 & 221\,724.256 & 0.395 &
-1.00(3)
                        \\
    5 &    2 &  E2-1 & &    4 &    2 &  E2-1 & 221\,728.700 & 0.396 &
-1.00(3)
                    \\
   10 &    2 &  B2 & &   10 &    1 &  B1 & 227\,545.019 & 8.759 & -1.15(3)
                    \\
    8 &    2 &  E1+1 & &    8 &    1 &  E1+1 & 227\,997.002 & 3.320 &
-1.00(3)
                    \\
    4 &    2 &  E1-1 & &    4 &    1 &  E1+1 & 229\,310.604 & 0.848 &
-0.83(3)
                    \\
    7 &    2 &  E2-1 & &    7 &    1 &  E2+1 & 229\,452.729 & 0.628 &
-0.96(3)
                    \\
    9 &    2 &  B1 & &    9 &    1 &  B2 & 231\,844.268 & 7.784 & -1.16(3)
                    \\
    5 &    2 &  E2+1 & &    5 &    1 &  E2+1 & 232\,003.755 & 3.580 &
-0.89(3)
                    \\
    7 &    2 &  A1 & &    7 &    1 &  A2 & 233\,368.424 & 5.922 & -1.03(3)
                    \\
\\[-1.8ex]
\hline\hline
\end{tabular}&

%%%%%%%%%%%%%%%%%%%%%%%%%%%%%%%%%%%%%%%%%%%%%%%%%%%%%%%%%%%%%%%%%%%%%%%%%%%%%%%%%%

\begin{tabular}[t]{ >{$}c<{$}>{$}c<{$}>{$}l<{$} c
 >{$}c<{$}>{$}c<{$}>{$}l<{$} D..{6} D..{6} D..{6}}
\hline\hline
\multicolumn{3}{c}{Upper state} & &
\multicolumn{3}{c}{Lower state} &
\multicolumn{1}{c}{Transition (MHz)} &
\multicolumn{1}{c}{S$\mu_{e}^{2}$ (D$^2$)} &
\multicolumn{1}{c}{$K_{\mu}$}\bigstrut\\\cline{1-3}\cline{5-7}
\\[-1.9ex]
J & K & \text{Sym} & & J & K & \text{Sym} & & &  \\
\hline
\\[-1.8ex]
   14 &    6 &  A1 & &   15 &    5 &  A2 & 235\,337.423 & 2.367 & -1.17(4)
                    \\
   14 &    6 &  A2 & &   15 &    5 &  A1 & 235\,337.540 & 2.367 & -1.17(4)
                    \\
    8 &    2 &  B2 & &    8 &    1 &  B1 & 235\,734.967 & 6.840 & -1.14(3)
                    \\
    6 &    2 &  A2 & &    6 &    1 &  A1 & 236\,408.779 & 5.020 & -1.03(3)
                    \\
    2 &    2 &  E1-1 & &    2 &    1 &  E1-1 & 237\,143.512 & 1.230 &
-0.88(3)
                    \\
    4 &    2 &  E1-1 & &    4 &    1 &  E1-1 & 239\,427.017 & 2.299 &
-0.87(3)
                    \\
    3 &    2 &  E1+1 & &    3 &    1 &  E1+1 & 239\,446.258 & 1.937 &
-0.98(3)
                    \\
    5 &    2 &  E1-1 & &    5 &    1 &  E1-1 & 241\,501.243 & 2.554 &
-0.87(3)
                    \\
    6 &    2 &  B2 & &    6 &    1 &  B1 & 242\,261.957 & 5.020 & -1.14(3)
                     \\
    6 &    2 &  E1-1 & &    6 &    1 &  E1-1 & 244\,151.624 & 2.725 &
-0.87(3)
                    \\
   10 &    5 &  B1 & &   11 &    4 &  B2 & 245\,463.443 & 1.506 & -1.09(3)
                    \\
   10 &    5 &  B2 & &   11 &    4 &  B1 & 245\,464.483 & 1.506 & -1.09(3)
                    \\
    2 &    2 &  A1 & &    2 &    1 &  A2 & 246\,924.172 & 1.298 & -1.03(3)
                     \\
    4 &    2 &  B2 & &    4 &    1 &  B1 & 247\,080.140 & 3.235 & -1.14(3)
                    \\
    7 &    2 &  E1-1 & &    7 &    1 &  E1-1 & 247\,362.353 & 2.807 &
-0.86(3)
                    \\
    3 &    2 &  B1 & &    3 &    1 &  B2 & 248\,838.499 & 2.317 & -1.14(3)
                    \\
    3 &    2 &  E2-1 & &    3 &    1 &  E2-1 & 248\,999.871 & 2.182 &
-1.09(3)
                    \\
    8 &    0 &  A1 & &    7 &    1 &  A2 & 250\,702.202 & 4.891 & -0.84(3)
                    \\
    6 &    2 &  E1+1 & &    6 &    1 &  E1-1 & 252\,908.786 & 1.740 &
-1.01(3)
                    \\
    6 &    2 &  E2-1 & &    6 &    1 &  E2-1 & 253\,768.569 & 3.999 &
-1.06(3)
                    \\
    4 &    1 &  E1-1 & &    3 &    0 &  E1+1 & 254\,055.766 & 0.259 &
-1.01(3)
                    \\
    9 &    2 &  E1-1 & &    9 &    1 &  E1-1 & 255\,444.689 & 2.612 &
-0.87(3)
                    \\
    4 &    2 &  B1 & &    4 &    1 &  B2 & 255\,997.777 & 3.065 & -1.13(3)
                    \\
    5 &    2 &  B2 & &    5 &    1 &  B1 & 258\,349.240 & 3.804 & -1.13(3)
                    \\
    7 &    2 &  A2 & &    7 &    1 &  A1 & 258\,857.426 & 5.080 & -1.03(3)
                    \\
   10 &    2 &  E1-1 & &   10 &    1 &  E1-1 & 260\,293.984 & 2.308 &
-0.87(3)
                    \\
   11 &    1 &  B2 & &   10 &    2 &  B1 & 260\,963.400 & 3.943 & -0.87(3)
                    \\
    4 &    1 &  E2+1 & &    3 &    0 &  E2+1 & 261\,024.312 & 3.128 &
-1.00(3)
                    \\
    4 &    1 &  B1 & &    3 &    0 &  B2 & 261\,219.282 & 3.924 & -0.96(3)
                    \\
    8 &    0 &  B1 & &    7 &    1 &  B2 & 261\,562.178 & 4.881 & -1.04(3)
                    \\
    8 &    0 &  E2+1 & &    7 &    1 &  E2+1 & 263\,377.814 & 4.613 &
-1.04(3)
                    \\
\\[-1.8ex]
\hline\hline
\end{tabular}\\
\end{tabular}
\end{table*}
\renewcommand{\arraystretch}{1.1}
\endgroup

\begingroup
\squeezetable
\renewcommand{\arraystretch}{1.1}
\begin{table*}
\caption{Transitions in methylamine (\ce{CH3NH2}) involving levels with an excitation 
energy lower than 10\,cm$^{-1}$. The fourth column lists the transition strength multiplied by 
the electric dipole moment, $\mu_{e}$, squared. The last column lists the sensitivity of the transitions
to a possible variation of the proton-to-electron mass ratio. The transitions labeled by roman numerals
correspond to the ones depicted in Fig.~\ref{Fig:levelscheme}. The transitions
labeled with an asterix have recently been detected by Muller \emph{et al.}~\cite{Muller2011}
in a cold cloud at $z=0.89$. \label{Tab:ColdTrans}}
\begin{tabular}[t]{ll}
\begin{tabular}[t]{>{$}r<{$} >{$}c<{$}>{$}c<{$}>{$}l<{$} c
 >{$}c<{$}>{$}c<{$}>{$}l<{$} D..{6} D..{6} D..{6}}
\cline{2-11}\\[-2.2ex]\cline{2-11}
&
\multicolumn{3}{c}{Upper state} & &
\multicolumn{3}{c}{Lower state} &
\multicolumn{1}{c}{Transition (MHz)} &
\multicolumn{1}{c}{S$\mu_{e}^{2}$ (D$^2$)} &
\multicolumn{1}{c}{$K_{\mu}$}\bigstrut\\\cline{2-4}\cline{6-8}
\\[-1.9ex]
& J & K & \text{Sym} & & J & K & \text{Sym} & & & \\
\cline{2-11}
\\[-1.8ex]
{\text{\sc i}}&1 &    1 &    A2 & &    1 &    1 &    A1 &    879.859 &
0.141&    -1.02(3)    \\
{\text{\sc ii}}&1 &    1 &    B2 & &    1 &    1 &    B1 &    881.386 &
0.142&    -1.02(3)    \\
{\text{\sc iii}}&1 &    1 &    A2 & &    2 &    0 &    A1 &
2\,166.305 & 0.779&    -19.1(6)    \\
&2 &    1 &    A1 & &    2 &    1 &    A2 &    2\,639.491 & 0.078&
-0.99(3)    \\
&2 &    1 &    B1 & &    2 &    1 &    B2 &    2\,644.073 & 0.080&
-0.98(3)    \\
{\text{\sc iv}}&2 &    0 &    E1+1 & &    1 &    1 &    E1+1 &
4\,364.348 & 0.456&    1.95(6)    \\
{\text{\sc v}}&1 &    1 &    E1+1 & &    1 &    1 &    E1-1 &
5\,094.897 & 0.004&    -4.0(1)    \\
&2 &    1 &    E1+1 & &    2 &    1 &    E1-1 &    5\,669.477 & 0.017&
-3.5(1)    \\
{\text{\sc vi}}&2 &    0 &    E2+1 & &    1 &    1 &    E2+1 &
6\,437.552 & 0.418&    -0.42(3)    \\
{\text{\sc vii}}&2 &    0 &    B1 & &    1 &    1 &    B2 &
8\,777.827 & 0.779&    -2.14(6)    \\
{\text{\sc viii}}&2 &    0 &    E1+1 & &    1 &    1 &    E1-1 &
9\,459.246 & 0.322&    -1.29(4)    \\
{\text{\sc ix}}&1 &    1 &    E2+1 & &    1 &    1 &    E2-1 &
11\,911.000 & 0.001&    -4.9(1)        \\
&2 &    1 &    E2+1 & &    2 &    1 &    E2-1 &    12\,167.419 & 0.004&
-4.8(1)    \\
{\text{\sc x}}&2 &    0 &    E2+1 & &    1 &    1 &    E2-1 &
18\,348.552 & 0.360&    -3.3(1)    \\
&3 &    0 &    A2 & &    2 &    1 &    A1 &    41\,263.780 & 1.541&
-0.05(3)    \\
&1 &    0 &    B2 & &    0 &    0 &    B1 &    44\,337.938 & 0.095&
-1.00(3)    \\
&1 &    0 &    E2+1 & &    0 &    0 &    E2+1 &    44\,338.468 & 0.094&
-1.00(3)    \\
&1 &    0 &    A2 & &    0 &    0 &    A1 &    44\,338.755 & 0.094&
-1.00(3)    \\
&1 &    0 &    E1+1 & &    0 &    0 &    E1+1 &    44\,338.876 & 0.094&
-1.00(3)    \\
&3 &    0 &    E1+1 & &    2 &    1 &    E1+1 &    48\,385.595 & 1.128&
-0.75(3)    \\
&3 &    0 &    E2+1 & &    2 &    1 &    E2+1 &    50\,615.856 & 0.936&
-0.94(3)    \\
&3 &    0 &    B2 & &    2 &    1 &    B1 &    52\,202.362 & 1.540&
-1.19(4)    \\
&3 &    0 &    E1+1 & &    2 &    1 &    E1-1 &    54\,055.072 & 0.412&
-1.04(3)    \\
&3 &    0 &    E2+1 & &    2 &    1 &    E2-1 &    62\,783.275 & 0.603&
-1.68(5)    \\
&2 &    1 &    E2-1 & &    2 &    0 &    E2+1 &    70\,199.113 & 2.420&
-0.40(3)    \\
&1 &    1 &    E2-1 & &    1 &    0 &    E2+1 &    70\,320.128 & 1.274&
-0.39(3) \\
&2 &    1 &    E2-1 & &    1 &    1 &    E2+1 &    76\,636.665 & 0.001&
-0.40(3)    \\
&2 &    1 &    B2 & &    2 &    0 &    B1 &    78\,135.504^* & 3.976&
-0.87(3)    \\
&2 &    1 &    E1-1 & &    2 &    0 &    E1+1 &    78\,928.726 & 2.914&
-0.98(3)    \\
&1 &    1 &    B1 & &    1 &    0 &    B2 &    79\,008.693^* & 2.373&
-0.87(3)    \\
&1 &    1 &    E1-1 & &    1 &    0 &    E1+1 &    79\,210.297 & 1.392&
-0.97(3)    \\
&1 &    1 &    E2+1 & &    1 &    0 &    E2+1 &    82\,231.128 & 1.099&
-1.05(3)    \\
&2 &    1 &    E2+1 & &    2 &    0 &    E2+1 &    82\,366.532 & 1.558&
-1.04(3)    \\
&2 &    1 &    E1-1 & &    1 &    1 &    E1+1 &    83\,293.074 & 0.003&
-0.82(3)    \\

\\[-1.8ex]
\cline{2-11}\\[-2.2ex]\cline{2-11}
\end{tabular}&

%%%%%%%%%%%%%%%%%%%%%%%%%%%%%%%%%%%%%%%%%%%%%%%%%%%%%%%%%%%%%%%%%%%%%%%%%%%%%%%%%%

\begin{tabular}[t]{ >{$}c<{$}>{$}c<{$}>{$}l<{$} c
 >{$}c<{$}>{$}c<{$}>{$}l<{$} D..{6} D..{6} D..{6}}
\cline{1-10}\\[-2.2ex]\cline{1-10}
\multicolumn{3}{c}{Upper state} & &
\multicolumn{3}{c}{Lower state} &
\multicolumn{1}{c}{Transition (MHz)} &
\multicolumn{1}{c}{S$\mu_{e}^{2}$ (D$^2$)} &
\multicolumn{1}{c}{$K_{\mu}$}\bigstrut\\\cline{1-3}\cline{5-7}
\\[-1.9ex]
J & K & \text{Sym} & & J & K & \text{Sym} & & &\\
\cline{1-10}
\\[-1.8ex]
1 &    1 &    E1+1 & &    1 &    0 &    E1+1 &    84\,305.195 & 0.982&
-1.15(3)    \\
2 &    1 &    E1+1 & &    2 &    0 &    E1+1 &    84\,598.202 & 1.065&
-1.14(3)    \\
2 &    1 &    B2 & &    1 &    1 &    B1 &    87\,794.717 & 0.141&
-1.00(3)    \\
2 &    1 &    A2 & &    1 &    1 &    A1 &    87\,795.016 & 0.141&
-1.00(3)    \\
2 &    1 &    E1-1 & &    1 &    1 &    E1-1 &    88\,387.971 & 0.138&
-1.01(3)    \\
2 &    1 &    E2-1 & &    1 &    1 &    E2-1 &    88\,547.665 & 0.140&
-1.01(3)    \\
2 &    0 &    B1 & &    1 &    0 &    B2 &    88\,667.906 & 0.189&
-1.00(3)    \\
2 &    0 &    E2+1 & &    1 &    0 &    E2+1 &    88\,668.681 & 0.189&
-1.00(3)    \\
2 &    0 &    E1+1 & &    1 &    0 &    E1+1 &    88\,669.543 & 0.188&
-1.00(3)    \\
2 &    0 &    A1 & &    1 &    0 &    A2 &    88\,669.626 & 0.188&
-1.00(3)    \\
2 &    1 &    E2+1 & &    1 &    1 &    E2+1 &    88\,804.084 & 0.141&
-0.99(3)    \\
2 &    1 &    E1+1 & &    1 &    1 &    E1+1 &    88\,962.550 & 0.138&
-0.99(3)    \\
2 &    1 &    A2 & &    2 &    0 &    A1 &    89\,081.463 & 3.978&
-1.44(4)    \\
2 &    1 &    A1 & &    1 &    1 &    A2 &    89\,554.649 & 0.141&
-1.00(3)    \\
2 &    1 &    B1 & &    1 &    1 &    B2 &    89\,557.404 & 0.141&
-1.00(3)    \\
1 &    1 &    A1 & &    1 &    0 &    A2 &    89\,956.072^* & 2.374&
-1.44(4)    \\
2 &    1 &    E1+1 & &    1 &    1 &    E1-1 &    94\,057.448 & 0.003&
-1.16(3)    \\
2 &    1 &    E2+1 & &    1 &    1 &    E2-1 &    100\,715.084 & 0.001&
-1.46(4)    \\
1 &    1 &    E2-1 & &    0 &    0 &    E2+1 &    114\,658.597 & 0.733&
-0.63(3)    \\
1 &    1 &    E1-1 & &    0 &    0 &    E1+1 &    123\,549.174 & 0.655&
-0.98(3)    \\
1 &    1 &    B2 & &    0 &    0 &    B1 &    124\,228.018 & 1.582&
-0.92(3)    \\
1 &    1 &    E2+1 & &    0 &    0 &    E2+1 &    126\,569.597 & 0.850&
-1.03(3)    \\
1 &    1 &    E1+1 & &    0 &    0 &    E1+1 &    128\,644.071 & 0.928&
-1.10(3)    \\
3 &    0 &    B2 & &    2 &    0 &    B1 &    132\,981.939 & 0.284&
-1.00(3)    \\
3 &    0 &    E2+1 & &    2 &    0 &    E2+1 &    132\,982.388 & 0.283&
-1.00(3)    \\
3 &    0 &    E1+1 & &    2 &    0 &    E1+1 &    132\,983.797 & 0.282&
-1.00(3)    \\
3 &    0 &    A2 & &    2 &    0 &    A1 &    132\,984.734 & 0.282&
-1.00(3)    \\
1 &    1 &    A2 & &    0 &    0 &    A1 &    135\,174.686 & 1.583&
-1.29(4)    \\
2 &    1 &    E2-1 & &    1 &    0 &    E2+1 &    158\,867.793 & 0.929&
-0.73(3)    \\
2 &    1 &    E1-1 & &    1 &    0 &    E1+1 &    167\,598.269 & 0.636&
-0.99(3)    \\
2 &    1 &    B1 & &    1 &    0 &    B2 &    169\,447.483 & 2.373&
-0.94(3)    \\
2 &    1 &    E2+1 & &    1 &    0 &    E2+1 &    171\,035.212 & 1.444&
-1.02(3)    \\
2 &    1 &    E1+1 & &    1 &    0 &    E1+1 &    173\,267.745 & 1.739&
-1.07(3)    \\
2 &    1 &    A1 & &    1 &    0 &    A2 &    180\,390.580 & 2.374&
-1.22(4)    \\

\\[-1.8ex]
\cline{1-10}\\[-2.2ex]\cline{1-10}
\end{tabular}\\
\end{tabular}
\end{table*}
\renewcommand{\arraystretch}{1.1}
\endgroup

\section{Sensitivity of selected transitions}
\label{Sec:Results}

Using the scaling relations for the high-barrier tunneling Hamiltonian determined in the previous section,
we are now able to calculate the sensitivity coefficient of any desired transition in the ground state of 
methylamine. In order to do numerical calculations, we rewrite Eq.~(\refeq{eq:Kmu}) as

%\begin{equation}
%K_{\mu}^{\nu_{mn}} = \frac{\nu_{mn}(\mu=\mu_{0}(1+\epsilon)) - \nu_{mn}(\mu=\mu_{0}(1-\epsilon))}
%{2\epsilon\nu_{mn}(\mu=\mu_{0})},
%\label{eq:Knumerically}
%\end{equation}

\begin{equation}
K_{\mu}^{\nu_{mn}} = 
\frac{\nu^{+}_{mn}-\nu^{-}_{mn}}{2\epsilon\nu_{mn}},
\label{eq:Knumerically}
\end{equation}

\noindent
with $\nu_{mn}$ the transition frequency between state $m$ and $n$ for the present value of $\mu$
and $\nu^{\pm}_{mn}$ the transition frequency when $\mu$ is replaced by $\mu(1\pm\epsilon)$ with $\epsilon$ a number
much smaller than 1 (in our calculations, we typically use $\epsilon=0.0001$). 
$\nu_{mn}$ is calculated using values for the molecular constants as listed in Table~\ref{Tab:MolecularConstants}, 
$\nu^{+}_{mn}$ and $\nu^{-}_{mn}$ are calculated using the molecular constants scaled according to the 
relations that were determined in the previous section.

We have calculated the $K_{\mu}$ coefficients for all rotational transitions in the ground state of methylamine 
with $J<30$ and $K_{a}<15$ and $\nu_{mn}$ below 500~GHz. The two largest coefficients $K_{\mu} \approx -19$ 
and $K_{\mu} \approx 24$  were found for the $1_{1} A2  \leftarrow 2_{0} A1$ and $13_{3} E1+1 \leftarrow 12_{4} E1+1$ 
transitions at 2166~MHz and 1458~MHz, respectively. 

In Table~\ref{Tab:Lovas}, the transitions of methylamine that are detected in 
astrophysical objects in our local galaxy are listed together with their transition strengths and 
sensitivity coefficients. Table~\ref{Tab:ColdTrans} lists 
transitions involving levels that have an excitation energy below 10~cm$^{-1}$, i.e., transitions involving
levels that are expected to be populated in cold molecular clouds. 
The rotational transitions labeled with an asterix have recently been detected by 
Muller \emph{et al.}~\cite{Muller2011} via absorption in a cold cloud at a redshift $z=0.89$. 
Due to their rather large transition frequency their sensitivity coefficients are only slightly enhanced.   
The transitions in Table~\ref{Tab:ColdTrans} that are labeled by the roman numerals {\sc i-x}, correspond to 
transitions in the $J=1,K=1$ and $J=2,K=0$ levels that are shown in the right hand side 
panel of Fig.~\ref{Fig:levelscheme}. The transitions labeled by {\sc i} and 
{\sc ii}, correspond to transitions between the levels of $K$-doublets, hence these have sensitivities 
of approximately $-1$. The transitions labeled by {\sc v} and {\sc ix} are transitions between levels 
which splittings are significantly affected by tunneling motions. The sensitivities of 
these transitions are on the order of $-5$, comparable to the sensitivity of the $h_{2v}$ and $h_{3v}$ parameters. 
The transitions labeled by {\sc iii,iv,vi,vii,viii} and {\sc x} are of particular interest as these are transitions
between levels that differ in overall rotational energy as well as torsional-wagging energy. 
Consequently, cancellation may take place that lead to an enhancement of the sensitivity coefficients. 
Of these, the transition labeled by {\sc iii} has the smallest transition frequency (2166~MHz) and the highest 
sensitivity coefficient ($K_{\mu}=-19$). The transition labeled by {\sc iv} at 4364~MHz has a sensitivity coefficient equal 
to $K_{\mu}=+2$. 
  
The estimated uncertainties of the $K_{\mu}$ coefficients 
are quoted in brackets in units of the last digits. There are two sources of the uncertainty in the $K_{\mu}$ 
coefficients: (i) the uncertainty in the determination of the molecular constants and (ii) inexactness of the 
scaling relations of the Hamiltonian parameters including errors due to neglecting the $\mu$ 
dependence of the torsion-wagging potential of the molecule. We have assumed the error in the scaling 
coefficients to be $\pm0.02$ for the rotational parameters, 
$\pm0.1$ for the tunneling parameters $h_{2v}$, $h_{3v}$, $q_{2}$, $r_{2}$ and $\pm1$ for higher order tunneling terms. 
Since the uncertainties for the measured transition frequencies in the ground torsional 
state of methylamine are less than $10^{-4}$ (and below 5$\times$10$^{-6}$ for the low-$J$ transitions 
of interest in the present study~\cite{Ilyushin2007}), we assume that the main errors in 
sensitivity coefficients are due to inexactness of the scaling relations of the Hamiltonian parameters. 
Therefore, similarly to the procedure adopted in Ref.~\cite{Levshakov2011}, the $K_{\mu}$ coefficients
were calculated taking either the upper or the lower bound for the scaling relations, corresponding to the upper and 
lower bounds of the assumed uncertainties. The difference was taken as an estimate 
of the uncertainty of the $K_{\mu}$ coefficients. In spite of the large uncertainties 
of the scaling relations for the higher order terms, the resulting errors in the
$K_{\mu}$ coefficients of the different transitions are below 3\%. To test the influence of the uncertainties in the scaling 
relations of the higher order terms, we have performed an additional calculation where only the non-tunneling 
parameters and $h_{2v}$, $h_{3v}$, $q_{2}$ and $r_{2}$ were used to calculate the $K_{\mu}$ coefficients for different 
transitions. The difference between this calculation and the calculation with the full set of scaling relations 
was less than 1.7\%, i.e., within the uncertainties presented in Table~\ref{Tab:Lovas} and \ref{Tab:ColdTrans}. 

It is interesting to note that almost identical values for the sensitivity coefficients are obtained 
by using an equation that directly connects the sensitivity coefficient of a transition with the
sensitivity coefficients of the Hamiltonian parameters:

\begin{equation}
K^{\nu_{mn}}_{\mu} = \frac{1}{\nu_{mn}}\sum_{s}K_{\mu}^{P_{s}}P_{s}
\left[\frac{\partial E_{n}}{\partial P_{s}} - \frac{\partial E_{m}}{\partial P_{s}}\right],
\label{eq:KHF}
\end{equation}

\noindent
where
\begin{equation}
\frac{\partial E_{m}}{\partial P_{s}} = \langle m\vert \hat{O}_{s} \vert m\rangle
\label{eq:EHF}
\end{equation}

\noindent
is the derivative of the energy level $E_{m}$ with respect to the Hamiltonian parameter $P_{s}$ 
used in the program to build up the least-squares-fit matrix, and $K_{\mu}^{P_{s}}$ is the sensitivity coefficient 
with respect to the $s$-th Hamiltonian parameter. Eq.~(\refeq{eq:KHF}) is based on the assumption that the 
energy of state $\vert m\rangle$ may be represented as $E_{m}= \sum_{s}P_{s}\langle m\vert \hat{O}_{s} \vert m\rangle$. 
This assumption is valid 
when the Hamiltonian depends linearly on the parameters, i.e. that the Hamiltonian may be written as $H = \sum_{s}P_{s}\hat{O}_{s}$. 
The high-barrier tunneling Hamiltonian used for methylamine depends nonlinearly on $\rho$, but as $K_{\mu}^{\rho}=0$, 
the transition sensitivity coefficients calculated using Eq.~(\refeq{eq:KHF}) agree well with 
the results obtained by using Eq.~(\refeq{eq:Knumerically}); the $\approx$\,0.4\% difference is attributed to 
the $\rho_{K}$ term, which is also non-linear and which scaling coefficient is nonzero.

From Eq.~(\refeq{eq:KHF}) it is seen that contributions to $K_{\mu}^{\nu_{mn}}$ from different terms 
in the Hamiltonian are proportional to the relative contributions of these terms to the transition frequency. 
From this fact, it is obvious that the resulting sensitivity coefficients are mainly determined by the 
largest terms in the Hamiltonian and uncertainties in the scaling relations for the high 
order parameters do not significantly affect our results.

Eq.~(\refeq{eq:KHF}), illustrates that the largest enhancement is obtained for transitions that 
connect two near degenerate levels that have substantially different dependences on $\mu$. 
The different dependence on $\mu$ is provided when the two levels 
contain non equal contributions from different types of motions in the molecule. In that case, 
a transition `converts' one superposition of rotation-torsion-wagging motion to another 
superposition of rotation-torsion-wagging motion. 
A significant enhancement is obtained when a `cancellation' takes place, i.e. when two levels have
nearly the same total energy due to quantitatively different contributions from various types of 
motion in the molecule.  

From Eq.~(\refeq{eq:KHF}), it is possible to obtain an upper limit for the sensitivity
coefficient that we may hope to find in the ground vibrational state of methylamine.
Considering the main, low order terms, the maximum splitting due to the
tunneling motions, i.e., the maximum torsional-wagging energy difference
between levels $n$ and $m$, may be roughly taken to be $4(h_{2v} + h_{3v})$. 
Large enhancements of the sensitivity are expected for transitions that convert
a considerable fraction of this energy into rotational energy. Using Eq.~(\refeq{eq:KHF})
and the values and sensitivities of the molecular parameters as listed in
Table~\ref{Tab:MolecularConstants}, the maximum sensitivity that we may hope to find is

\begin{align}
K_{\mu} &= K_{\mu}^{\rm rot} \pm \frac{1}{\nu_{nm}}\left( 4h_{2v} \left[K_{\mu}^{h_{2v}}-K_{\mu}^{\rm rot}\right] \right. \nonumber\\
&~~~~~~~~~~~~~~~~~~+ \left. 4h_{3v} \left[K_{\mu}^{h_{3v}}-K_{\mu}^{\rm rot} \right]\right)  \nonumber\\
&\approx -1 \pm 64\,800/\nu_{nm} ,
\label{eq:K7}
\end{align} 

\noindent
with $K_{\mu}^{\rm rot} =-1$ (i.e. that the $K_{\mu}$ of a rotational parameters) 
and $\nu_{nm}$ the transition frequency in MHz. 
The sensitivities obtained from our numerical calculations are indeed found within these bounds.

\section{Conclusion}

Spectra of molecular hydrogen in highly redshifted objects have been used to constrain
a possible variation of the proton-electron mass ratio $\mu$ since the 1970s.
However, as the observed absorptions in H$_2$ correspond to transitions between different electronic states,
these are rather insensitive to $\mu$; the sensitivity coefficients $K_{\mu}$ 
are in the range $(-0.01, +0.05)$~\cite{Thomphson1975,Varshalovich1993,Ubachs2007}. 
For this reason even the highest quality \ce{H2} 
absorption spectra involving over 90 lines, observed with the large dish Keck 
Telescope~\cite{Malec2010} and the Very Large Telescope~\cite{Weerdenburg2011} yield constraints 
$|\Delta\mu/\mu|$ at the level of only $5 \times 10^{-6}$.

The notion that specific molecules exhibit an enhanced sensitivity to $\mu$ variation is changing 
the paradigm for searching drifting constants on a cosmological time scales from the optical to the 
radio domain. The use of the \ce{NH_3} inversion transitions in the microwave range that have
$K_{\mu}$-coefficients of $-4.2$, has led to much tighter constraints on 
$\Delta\mu/\mu$~\cite{Veldhoven2004,FlambaumNH3_2007,Murphy2008,Kanekar2011}. 
It was recently pointed out that microwave transitions in the methanol molecule (\ce{CH$_3$OH}) 
have sensitivity coefficients in the range $(-42, +53)$~\cite{Jansen2011PRL,Levshakov2011}.
In this paper, we showed that the sensitivity of microwave transitions in methylamine, \ce{CH_{3}NH_{2}}, 
are in the range $(-19,+24)$. 

Methylamine is particularly relevant as it was recently observed at $z = 0.8859$ in the intervening galaxy 
towards the quasar PKS 1830--211~\cite{Muller2011}.
The sensitivity coefficients of the observed transitions at 78.135, 79.008, and 89.956\,GHz transitions 
were calculated to be $K_\mu = -0.87$ for the first two and $K_\mu = -1.4$ 
for the third transition, respectively (see Table~\ref{Tab:ColdTrans}). These three methylamine lines 
have a mean radial velocity of $v_{\ce{CH3NH2}} = -6.2\pm1.6$ km~s$^{-1}$
~\cite{Muller2011}. With $|\Delta K_\mu| = 0.563$ and the uncertainty interval $\Delta v = 1.6$ km~s$^{-1}$,
we obtain a preliminary estimate of $\Delta \mu/\mu$:

\begin{equation}
\left| \frac{\Delta \mu}{\mu} \right| = 
\left| \frac{\Delta v}{c\Delta K_\mu} \right| < 9\times10^{-6}\,,
\end{equation}

\noindent
where $c$ is the speed of light.

A tighter constraint on $\Delta \mu/\mu$ is obtained from the comparison of 
$v_{\ce{CH3NH2}}$ with the radial velocity of the methanol line at 60.531\,GHz 
also detected at $z=0.8859$; $v_{\ce{CH3OH}} = -5.3\pm0.5$ km~s$^{-1}$~\cite{Muller2011}.
According to Ref.~\cite{Jansen2011PRL}, this transition has a sensitivity coefficient $K_\mu=-7.4$. 
In this case we have $|\Delta K_\mu|=6.5$ and $\Delta v=0.9\pm1.7$ km~s$^{-1}$, which yields
$|\Delta \mu/\mu|< 10^{-6}$. This estimate contains an unknown input due to possible 
non-co-spacial distribution of \ce{CH3OH} and \ce{CH3NH2}. More robust constraints
on $\Delta \mu/\mu$ are derived from observations of lines of the same molecule. 
In this approach the low frequency transitions of \ce{CH3NH2} at 2166 and
4364\,MHz would be particularly attractive as the difference of their sensitivity
coefficients is $\Delta K_\mu \approx 21$.

\section{Acknowledgments}

This research has been supported by NWO via a VIDI-grant and by the ERC via a Starting Grant.
M.G.K. and S.A.L. are supported in part by the DFG grant SFB 676 Teilprojekt C4 and by
the RFBR grant 11-02-12284-ofi-m-2011. W.U. acknowledges support from the Netherlands Foundation 
for the Research of Matter (FOM). We thank Jon Hougen for his continuing interest for this project 
and invaluable help.


\begin{thebibliography}{34}%
\makeatletter
\providecommand \@ifxundefined [1]{%
 \@ifx{#1\undefined}
}%
\providecommand \@ifnum [1]{%
 \ifnum #1\expandafter \@firstoftwo
 \else \expandafter \@secondoftwo
 \fi
}%
\providecommand \@ifx [1]{%
 \ifx #1\expandafter \@firstoftwo
 \else \expandafter \@secondoftwo
 \fi
}%
\providecommand \natexlab [1]{#1}%
\providecommand \enquote  [1]{``#1''}%
\providecommand \bibnamefont  [1]{#1}%
\providecommand \bibfnamefont [1]{#1}%
\providecommand \citenamefont [1]{#1}%
\providecommand \href@noop [0]{\@secondoftwo}%
\providecommand \href [0]{\begingroup \@sanitize@url \@href}%
\providecommand \@href[1]{\@@startlink{#1}\@@href}%
\providecommand \@@href[1]{\endgroup#1\@@endlink}%
\providecommand \@sanitize@url [0]{\catcode `\\12\catcode `\$12\catcode
  `\&12\catcode `\#12\catcode `\^12\catcode `\_12\catcode `\%12\relax}%
\providecommand \@@startlink[1]{}%
\providecommand \@@endlink[0]{}%
\providecommand \url  [0]{\begingroup\@sanitize@url \@url }%
\providecommand \@url [1]{\endgroup\@href {#1}{\urlprefix }}%
\providecommand \urlprefix  [0]{URL }%
\providecommand \Eprint [0]{\href }%
\providecommand \doibase [0]{http://dx.doi.org/}%
\providecommand \selectlanguage [0]{\@gobble}%
\providecommand \bibinfo  [0]{\@secondoftwo}%
\providecommand \bibfield  [0]{\@secondoftwo}%
\providecommand \translation [1]{[#1]}%
\providecommand \BibitemOpen [0]{}%
\providecommand \bibitemStop [0]{}%
\providecommand \bibitemNoStop [0]{.\EOS\space}%
\providecommand \EOS [0]{\spacefactor3000\relax}%
\providecommand \BibitemShut  [1]{\csname bibitem#1\endcsname}%
\let\auto@bib@innerbib\@empty
%</preamble>
\bibitem [{\citenamefont {Kozlov}\ \emph {et~al.}(2011)\citenamefont {Kozlov},
  \citenamefont {Porsev},\ and\ \citenamefont {Reimers}}]{Kozlov2011PRA}%
  \BibitemOpen
  \bibfield  {author} {\bibinfo {author} {\bibfnamefont {M.~G.}\ \bibnamefont
  {Kozlov}}, \bibinfo {author} {\bibfnamefont {S.~G.}\ \bibnamefont {Porsev}},
  \ and\ \bibinfo {author} {\bibfnamefont {D.}~\bibnamefont {Reimers}},\
  }\href@noop {} {\bibfield  {journal} {\bibinfo  {journal} {Phys. Rev. A}\
  }\textbf {\bibinfo {volume} {83}},\ \bibinfo {pages} {052123} (\bibinfo
  {year} {2011})}\BibitemShut {NoStop}%
\bibitem [{Note1()}]{Note1}%
  \BibitemOpen
  \bibinfo {note} {In Kozlov~\protect \emph {et al.}~\cite {Kozlov2011PRA},
  $\mu $ is defined as the electron-to-proton mass ratio, consequently, the
  sensitivity coefficient used in that work is minus times the sensitivity
  coefficient used here, $Q_{\mu }$=$-K_{\mu }$.}\BibitemShut {Stop}%
\bibitem [{\citenamefont {Jansen}\ \emph
  {et~al.}(2011{\natexlab{a}})\citenamefont {Jansen}, \citenamefont {Xu},
  \citenamefont {Kleiner}, \citenamefont {Ubachs},\ and\ \citenamefont
  {Bethlem}}]{Jansen2011PRL}%
  \BibitemOpen
  \bibfield  {author} {\bibinfo {author} {\bibfnamefont {P.}~\bibnamefont
  {Jansen}}, \bibinfo {author} {\bibfnamefont {L.-H.}\ \bibnamefont {Xu}},
  \bibinfo {author} {\bibfnamefont {I.}~\bibnamefont {Kleiner}}, \bibinfo
  {author} {\bibfnamefont {W.}~\bibnamefont {Ubachs}}, \ and\ \bibinfo {author}
  {\bibfnamefont {H.~L.}\ \bibnamefont {Bethlem}},\ }\href {\doibase
  10.1103/PhysRevLett.106.100801} {\bibfield  {journal} {\bibinfo  {journal}
  {Phys. Rev. Lett.}\ }\textbf {\bibinfo {volume} {106}},\ \bibinfo {pages}
  {100801} (\bibinfo {year} {2011}{\natexlab{a}})}\BibitemShut {NoStop}%
\bibitem [{\citenamefont {Jansen}\ \emph
  {et~al.}(2011{\natexlab{b}})\citenamefont {Jansen}, \citenamefont {Kleiner},
  \citenamefont {Xu}, \citenamefont {Ubachs},\ and\ \citenamefont
  {Bethlem}}]{Jansen2011PRA}%
  \BibitemOpen
  \bibfield  {author} {\bibinfo {author} {\bibfnamefont {P.}~\bibnamefont
  {Jansen}}, \bibinfo {author} {\bibfnamefont {I.}~\bibnamefont {Kleiner}},
  \bibinfo {author} {\bibfnamefont {L.-H.}\ \bibnamefont {Xu}}, \bibinfo
  {author} {\bibfnamefont {W.}~\bibnamefont {Ubachs}}, \ and\ \bibinfo {author}
  {\bibfnamefont {H.~L.}\ \bibnamefont {Bethlem}},\ }\href@noop {} {\bibfield
  {journal} {\bibinfo  {journal} {Phys. Rev. A}\ }\textbf {\bibinfo {volume}
  {84}},\ \bibinfo {pages} {062505} (\bibinfo {year}
  {2011}{\natexlab{b}})}\BibitemShut {NoStop}%
\bibitem [{\citenamefont {Levshakov}\ \emph {et~al.}(2011)\citenamefont
  {Levshakov}, \citenamefont {Kozlov},\ and\ \citenamefont
  {Reimers}}]{Levshakov2011}%
  \BibitemOpen
  \bibfield  {author} {\bibinfo {author} {\bibfnamefont {S.~A.}\ \bibnamefont
  {Levshakov}}, \bibinfo {author} {\bibfnamefont {M.~G.}\ \bibnamefont
  {Kozlov}}, \ and\ \bibinfo {author} {\bibfnamefont {D.}~\bibnamefont
  {Reimers}},\ }\href {http://stacks.iop.org/0004-637X/738/i=1/a=26} {\bibfield
   {journal} {\bibinfo  {journal} {Astrophys. J.}\ }\textbf {\bibinfo {volume}
  {738}},\ \bibinfo {pages} {26} (\bibinfo {year} {2011})}\BibitemShut
  {NoStop}%
\bibitem [{\citenamefont {Muller}\ \emph {et~al.}(2011)\citenamefont {Muller},
  \citenamefont {Beelen}, \citenamefont {Gu\'elin}, \citenamefont {Aalto},
  \citenamefont {Black}, \citenamefont {Combes}, \citenamefont {Curran},
  \citenamefont {Theule},\ and\ \citenamefont {Longmore}}]{Muller2011}%
  \BibitemOpen
  \bibfield  {author} {\bibinfo {author} {\bibfnamefont {S.}~\bibnamefont
  {Muller}}, \bibinfo {author} {\bibfnamefont {A.}~\bibnamefont {Beelen}},
  \bibinfo {author} {\bibfnamefont {M.}~\bibnamefont {Gu\'elin}}, \bibinfo
  {author} {\bibfnamefont {S.}~\bibnamefont {Aalto}}, \bibinfo {author}
  {\bibfnamefont {J.~H.}\ \bibnamefont {Black}}, \bibinfo {author}
  {\bibfnamefont {F.}~\bibnamefont {Combes}}, \bibinfo {author} {\bibfnamefont
  {S.}~\bibnamefont {Curran}}, \bibinfo {author} {\bibfnamefont
  {P.}~\bibnamefont {Theule}}, \ and\ \bibinfo {author} {\bibfnamefont
  {S.}~\bibnamefont {Longmore}},\ }\href
  {http://dx.doi.org/10.1051/0004-6361/201117096} {\bibfield  {journal}
  {\bibinfo  {journal} {Astron. Astrophys.}\ }\textbf {\bibinfo {volume}
  {535}},\ \bibinfo {pages} {A103} (\bibinfo {year} {2011})}\BibitemShut
  {NoStop}%
\bibitem [{\citenamefont {Ilyushin}\ \emph {et~al.}(2010)\citenamefont
  {Ilyushin}, \citenamefont {Cloessner}, \citenamefont {Chou}, \citenamefont
  {Picraux}, \citenamefont {Hougen},\ and\ \citenamefont
  {Lavrich}}]{Ilyushin_JCP2010}%
  \BibitemOpen
  \bibfield  {author} {\bibinfo {author} {\bibfnamefont {V.~V.}\ \bibnamefont
  {Ilyushin}}, \bibinfo {author} {\bibfnamefont {E.~A.}\ \bibnamefont
  {Cloessner}}, \bibinfo {author} {\bibfnamefont {Y.-C.}\ \bibnamefont {Chou}},
  \bibinfo {author} {\bibfnamefont {L.~B.}\ \bibnamefont {Picraux}}, \bibinfo
  {author} {\bibfnamefont {J.~T.}\ \bibnamefont {Hougen}}, \ and\ \bibinfo
  {author} {\bibfnamefont {R.}~\bibnamefont {Lavrich}},\ }\href@noop {}
  {\bibfield  {journal} {\bibinfo  {journal} {J. Chem. Phys.}\ }\textbf
  {\bibinfo {volume} {133}},\ \bibinfo {pages} {184307} (\bibinfo {year}
  {2010})}\BibitemShut {NoStop}%
\bibitem [{\citenamefont {Ohashi}\ and\ \citenamefont
  {Hougen}(1987)}]{Ohashi&Hougen1987}%
  \BibitemOpen
  \bibfield  {author} {\bibinfo {author} {\bibfnamefont {N.}~\bibnamefont
  {Ohashi}}\ and\ \bibinfo {author} {\bibfnamefont {J.~T.}\ \bibnamefont
  {Hougen}},\ }\href@noop {} {\bibfield  {journal} {\bibinfo  {journal} {J.
  Mol. Spectrosc.}\ }\textbf {\bibinfo {volume} {121}},\ \bibinfo {pages} {474}
  (\bibinfo {year} {1987})}\BibitemShut {NoStop}%
\bibitem [{\citenamefont {Ilyushin}\ \emph {et~al.}(2005)\citenamefont
  {Ilyushin}, \citenamefont {Alekseev}, \citenamefont {Dyubko}, \citenamefont
  {Motiyenko},\ and\ \citenamefont {Hougen}}]{Ilyushin2005}%
  \BibitemOpen
  \bibfield  {author} {\bibinfo {author} {\bibfnamefont {V.~V.}\ \bibnamefont
  {Ilyushin}}, \bibinfo {author} {\bibfnamefont {E.~A.}\ \bibnamefont
  {Alekseev}}, \bibinfo {author} {\bibfnamefont {S.~F.}\ \bibnamefont
  {Dyubko}}, \bibinfo {author} {\bibfnamefont {R.~A.}\ \bibnamefont
  {Motiyenko}}, \ and\ \bibinfo {author} {\bibfnamefont {J.~T.}\ \bibnamefont
  {Hougen}},\ }\href {\doibase 10.1016/j.jms.2004.08.022} {\bibfield  {journal}
  {\bibinfo  {journal} {J. Mol. Spectrosc.}\ }\textbf {\bibinfo {volume}
  {229}},\ \bibinfo {pages} {170 } (\bibinfo {year} {2005})}\BibitemShut
  {NoStop}%
\bibitem [{\citenamefont {Ilyushin}\ and\ \citenamefont
  {Lovas}(2007)}]{Ilyushin2007}%
  \BibitemOpen
  \bibfield  {author} {\bibinfo {author} {\bibfnamefont {V.~V.}\ \bibnamefont
  {Ilyushin}}\ and\ \bibinfo {author} {\bibfnamefont {F.~J.}\ \bibnamefont
  {Lovas}},\ }\href {\doibase DOI:10.1063/1.2769382} {\bibfield  {journal}
  {\bibinfo  {journal} {J. Phys. Chem. Rev. Data}\ }\textbf {\bibinfo {volume}
  {36}},\ \bibinfo {pages} {1141} (\bibinfo {year} {2007})}\BibitemShut
  {NoStop}%
\bibitem [{\citenamefont {Hougen}\ and\ \citenamefont
  {DeKoven}(1983)}]{Hougen&DeKoven1983}%
  \BibitemOpen
  \bibfield  {author} {\bibinfo {author} {\bibfnamefont {J.~T.}\ \bibnamefont
  {Hougen}}\ and\ \bibinfo {author} {\bibfnamefont {B.~M.}\ \bibnamefont
  {DeKoven}},\ }\href {\doibase 10.1016/0022-2852(83)90249-7} {\bibfield
  {journal} {\bibinfo  {journal} {J. Mol. Spectrosc.}\ }\textbf {\bibinfo
  {volume} {98}},\ \bibinfo {pages} {375 } (\bibinfo {year}
  {1983})}\BibitemShut {NoStop}%
\bibitem [{onl()}]{onlinematerial}%
  \BibitemOpen
  \href@noop {} {}\bibinfo {note} {See EPAPS Document No. ??? For more
  information on EPAPS, see
  \url{http://www.aip.org/pubservs/epaps.html}.}\BibitemShut {Stop}%
\bibitem [{\citenamefont {Ohashi}\ and\ \citenamefont
  {Hougen}(1985)}]{Ohashi&Hougen1985}%
  \BibitemOpen
  \bibfield  {author} {\bibinfo {author} {\bibfnamefont {N.}~\bibnamefont
  {Ohashi}}\ and\ \bibinfo {author} {\bibfnamefont {J.~T.}\ \bibnamefont
  {Hougen}},\ }\href {\doibase 10.1016/0022-2852(85)90170-5} {\bibfield
  {journal} {\bibinfo  {journal} {J. Mol. Spectrosc.}\ }\textbf {\bibinfo
  {volume} {112}},\ \bibinfo {pages} {384 } (\bibinfo {year}
  {1985})}\BibitemShut {NoStop}%
\bibitem [{\citenamefont {Takagi}\ and\ \citenamefont
  {Kojima}(1971)}]{Takagi1971}%
  \BibitemOpen
  \bibfield  {author} {\bibinfo {author} {\bibfnamefont {K.}~\bibnamefont
  {Takagi}}\ and\ \bibinfo {author} {\bibfnamefont {T.}~\bibnamefont
  {Kojima}},\ }\href@noop {} {\bibfield  {journal} {\bibinfo  {journal} {J.
  Phys. Soc. Jpn.}\ }\textbf {\bibinfo {volume} {30}},\ \bibinfo {pages} {1145}
  (\bibinfo {year} {1971})}\BibitemShut {NoStop}%
\bibitem [{\citenamefont {Kr\'eglewski}\ \emph
  {et~al.}(1990{\natexlab{a}})\citenamefont {Kr\'eglewski}, \citenamefont
  {Jager},\ and\ \citenamefont {Dreizler}}]{Kreglewski1990CD3NH2}%
  \BibitemOpen
  \bibfield  {author} {\bibinfo {author} {\bibfnamefont {M.}~\bibnamefont
  {Kr\'eglewski}}, \bibinfo {author} {\bibfnamefont {W.}~\bibnamefont {Jager}},
  \ and\ \bibinfo {author} {\bibfnamefont {H.}~\bibnamefont {Dreizler}},\
  }\href@noop {} {\bibfield  {journal} {\bibinfo  {journal} {J. Mol.
  Spectrosc.}\ }\textbf {\bibinfo {volume} {144}},\ \bibinfo {pages} {334}
  (\bibinfo {year} {1990}{\natexlab{a}})}\BibitemShut {NoStop}%
\bibitem [{\citenamefont {Kr\'eglewski}\ \emph
  {et~al.}(1990{\natexlab{b}})\citenamefont {Kr\'eglewski}, \citenamefont
  {Stryjewski},\ and\ \citenamefont {Dreizler}}]{Kreglewski1990CD3ND2}%
  \BibitemOpen
  \bibfield  {author} {\bibinfo {author} {\bibfnamefont {M.}~\bibnamefont
  {Kr\'eglewski}}, \bibinfo {author} {\bibfnamefont {D.}~\bibnamefont
  {Stryjewski}}, \ and\ \bibinfo {author} {\bibfnamefont {H.}~\bibnamefont
  {Dreizler}},\ }\href@noop {} {\bibfield  {journal} {\bibinfo  {journal} {J.
  Mol. Spectrosc.}\ }\textbf {\bibinfo {volume} {139}},\ \bibinfo {pages} {182}
  (\bibinfo {year} {1990}{\natexlab{b}})}\BibitemShut {NoStop}%
\bibitem [{\citenamefont {Ilyushin~\emph{et al.}}()}]{Ilyushin:inpreparation}%
  \BibitemOpen
  \bibfield  {author} {\bibinfo {author} {\bibfnamefont {V.~V.}\ \bibnamefont
  {Ilyushin~\emph{et al.}}},\ }\href@noop {} {\bibinfo  {journal} {in
  preparation}\ }\BibitemShut {NoStop}%
\bibitem [{\citenamefont {Smeyers}\ \emph {et~al.}(1996)\citenamefont
  {Smeyers}, \citenamefont {Villa},\ and\ \citenamefont
  {Senent}}]{Smeyers1996}%
  \BibitemOpen
\bibfield  {journal} {  }\bibfield  {author} {\bibinfo {author} {\bibfnamefont
  {Y.~G.}\ \bibnamefont {Smeyers}}, \bibinfo {author} {\bibfnamefont
  {M.}~\bibnamefont {Villa}}, \ and\ \bibinfo {author} {\bibfnamefont {M.~L.}\
  \bibnamefont {Senent}},\ }\href@noop {} {\bibfield  {journal} {\bibinfo
  {journal} {J. Mol. Spectrosc.}\ }\textbf {\bibinfo {volume} {177}},\ \bibinfo
  {pages} {66} (\bibinfo {year} {1996})}\BibitemShut {NoStop}%
\bibitem [{\citenamefont {Kr\'eglewski}\ and\ \citenamefont
  {Wlodarczak}(1992)}]{Kreglewski1992}%
  \BibitemOpen
  \bibfield  {author} {\bibinfo {author} {\bibfnamefont {M.}~\bibnamefont
  {Kr\'eglewski}}\ and\ \bibinfo {author} {\bibfnamefont {G.}~\bibnamefont
  {Wlodarczak}},\ }\href {\doibase 10.1016/0022-2852(92)90239-K} {\bibfield
  {journal} {\bibinfo  {journal} {J. Mol. Spectrosc.}\ }\textbf {\bibinfo
  {volume} {156}},\ \bibinfo {pages} {383 } (\bibinfo {year}
  {1992})}\BibitemShut {NoStop}%
\bibitem [{\citenamefont {Smeyers}\ \emph {et~al.}(1998)\citenamefont
  {Smeyers}, \citenamefont {Villa},\ and\ \citenamefont
  {Senent}}]{Smeyers1998}%
  \BibitemOpen
  \bibfield  {author} {\bibinfo {author} {\bibfnamefont {Y.~G.}\ \bibnamefont
  {Smeyers}}, \bibinfo {author} {\bibfnamefont {M.}~\bibnamefont {Villa}}, \
  and\ \bibinfo {author} {\bibfnamefont {M.~L.}\ \bibnamefont {Senent}},\
  }\href@noop {} {\bibfield  {journal} {\bibinfo  {journal} {J. Mol.
  Spectrosc.}\ }\textbf {\bibinfo {volume} {191}},\ \bibinfo {pages} {232}
  (\bibinfo {year} {1998})}\BibitemShut {NoStop}%
\bibitem [{\citenamefont {Tsuboi}\ \emph {et~al.}(1966)\citenamefont {Tsuboi},
  \citenamefont {Hirakawa},\ and\ \citenamefont {Tamagake}}]{Tsuboi1966}%
  \BibitemOpen
  \bibfield  {author} {\bibinfo {author} {\bibfnamefont {M.}~\bibnamefont
  {Tsuboi}}, \bibinfo {author} {\bibfnamefont {A.~Y.}\ \bibnamefont
  {Hirakawa}}, \ and\ \bibinfo {author} {\bibfnamefont {K.}~\bibnamefont
  {Tamagake}},\ }\href@noop {} {\bibfield  {journal} {\bibinfo  {journal}
  {Proc. Japan Acad.}\ }\textbf {\bibinfo {volume} {42}},\ \bibinfo {pages}
  {795} (\bibinfo {year} {1966})}\BibitemShut {NoStop}%
\bibitem [{\citenamefont {Sztraka}(1987)}]{Sztraka1987}%
  \BibitemOpen
  \bibfield  {author} {\bibinfo {author} {\bibfnamefont {L.}~\bibnamefont
  {Sztraka}},\ }\href@noop {} {\bibfield  {journal} {\bibinfo  {journal} {Acta
  Chim. Hung.}\ }\textbf {\bibinfo {volume} {6}},\ \bibinfo {pages} {865}
  (\bibinfo {year} {1987})}\BibitemShut {NoStop}%
\bibitem [{\citenamefont {Ohashi}\ \emph {et~al.}(1988)\citenamefont {Ohashi},
  \citenamefont {Takagi}, \citenamefont {Hougen}, \citenamefont {Olson},\ and\
  \citenamefont {Lafferty}}]{Ohashi1988}%
  \BibitemOpen
  \bibfield  {author} {\bibinfo {author} {\bibfnamefont {N.}~\bibnamefont
  {Ohashi}}, \bibinfo {author} {\bibfnamefont {K.}~\bibnamefont {Takagi}},
  \bibinfo {author} {\bibfnamefont {J.~T.}\ \bibnamefont {Hougen}}, \bibinfo
  {author} {\bibfnamefont {W.~B.}\ \bibnamefont {Olson}}, \ and\ \bibinfo
  {author} {\bibfnamefont {W.~J.}\ \bibnamefont {Lafferty}},\ }\href@noop {}
  {\bibfield  {journal} {\bibinfo  {journal} {J. Mol. Spectrosc.}\ }\textbf
  {\bibinfo {volume} {132}},\ \bibinfo {pages} {242} (\bibinfo {year}
  {1988})}\BibitemShut {NoStop}%
\bibitem [{\citenamefont {Tsuboi}\ \emph {et~al.}(1964)\citenamefont {Tsuboi},
  \citenamefont {Hirakawa}, \citenamefont {Ino}, \citenamefont {Sasaki},\ and\
  \citenamefont {Tamagake}}]{Tsuboi1964}%
  \BibitemOpen
  \bibfield  {author} {\bibinfo {author} {\bibfnamefont {M.}~\bibnamefont
  {Tsuboi}}, \bibinfo {author} {\bibfnamefont {A.~Y.}\ \bibnamefont
  {Hirakawa}}, \bibinfo {author} {\bibfnamefont {T.}~\bibnamefont {Ino}},
  \bibinfo {author} {\bibfnamefont {T.}~\bibnamefont {Sasaki}}, \ and\ \bibinfo
  {author} {\bibfnamefont {K.}~\bibnamefont {Tamagake}},\ }\href {\doibase
  DOI:10.1063/1.1726344} {\bibfield  {journal} {\bibinfo  {journal} {J. Chem.
  Phys.}\ }\textbf {\bibinfo {volume} {41}},\ \bibinfo {pages} {2721} (\bibinfo
  {year} {1964})}\BibitemShut {NoStop}%
\bibitem [{\citenamefont {Lovas}(2004)}]{Lovas2004}%
  \BibitemOpen
  \bibfield  {author} {\bibinfo {author} {\bibfnamefont {F.~J.}\ \bibnamefont
  {Lovas}},\ }\href {\doibase 10.1063/1.1633275} {\bibfield  {journal}
  {\bibinfo  {journal} {J. Phys. Chem. Ref. Data}\ }\textbf {\bibinfo {volume}
  {33}},\ \bibinfo {pages} {177} (\bibinfo {year} {2004})}\BibitemShut
  {NoStop}%
\bibitem [{\citenamefont {Thomphson}(1975)}]{Thomphson1975}%
  \BibitemOpen
  \bibfield  {author} {\bibinfo {author} {\bibfnamefont {R.~I.}\ \bibnamefont
  {Thomphson}},\ }\href {h} {\bibfield  {journal} {\bibinfo  {journal}
  {Astrophys. Lett.}\ }\textbf {\bibinfo {volume} {16}} (\bibinfo {year}
  {1975})}\BibitemShut {NoStop}%
\bibitem [{\citenamefont {Varshalovich}\ and\ \citenamefont
  {Levshakov}(1993)}]{Varshalovich1993}%
  \BibitemOpen
  \bibfield  {author} {\bibinfo {author} {\bibfnamefont {D.~A.}\ \bibnamefont
  {Varshalovich}}\ and\ \bibinfo {author} {\bibfnamefont {S.~A.}\ \bibnamefont
  {Levshakov}},\ }\href
  {http://www.jetpletters.ac.ru/ps/1187/article_17908.shtml} {\bibfield
  {journal} {\bibinfo  {journal} {JETP Lett.}\ }\textbf {\bibinfo {volume}
  {58}},\ \bibinfo {pages} {231} (\bibinfo {year} {1993})}\BibitemShut
  {NoStop}%
\bibitem [{\citenamefont {Ubachs}\ \emph {et~al.}(2007)\citenamefont {Ubachs},
  \citenamefont {Buning}, \citenamefont {Eikema},\ and\ \citenamefont
  {Reinhold}}]{Ubachs2007}%
  \BibitemOpen
  \bibfield  {author} {\bibinfo {author} {\bibfnamefont {W.}~\bibnamefont
  {Ubachs}}, \bibinfo {author} {\bibfnamefont {R.}~\bibnamefont {Buning}},
  \bibinfo {author} {\bibfnamefont {K.}~\bibnamefont {Eikema}}, \ and\ \bibinfo
  {author} {\bibfnamefont {E.}~\bibnamefont {Reinhold}},\ }\href {\doibase DOI:
  10.1016/j.jms.2006.12.004} {\bibfield  {journal} {\bibinfo  {journal} {J.
  Mol. Spectrosc.}\ }\textbf {\bibinfo {volume} {241}},\ \bibinfo {pages} {155
  } (\bibinfo {year} {2007})}\BibitemShut {NoStop}%
\bibitem [{\citenamefont {Malec}\ \emph {et~al.}(2010)\citenamefont {Malec},
  \citenamefont {Buning}, \citenamefont {Murphy}, \citenamefont {Milutinovic},
  \citenamefont {Ellison}, \citenamefont {Prochaska}, \citenamefont {Kaper},
  \citenamefont {Tumlinson}, \citenamefont {Carswell},\ and\ \citenamefont
  {Ubachs}}]{Malec2010}%
  \BibitemOpen
  \bibfield  {author} {\bibinfo {author} {\bibfnamefont {A.~L.}\ \bibnamefont
  {Malec}}, \bibinfo {author} {\bibfnamefont {R.}~\bibnamefont {Buning}},
  \bibinfo {author} {\bibfnamefont {M.~T.}\ \bibnamefont {Murphy}}, \bibinfo
  {author} {\bibfnamefont {N.}~\bibnamefont {Milutinovic}}, \bibinfo {author}
  {\bibfnamefont {S.~L.}\ \bibnamefont {Ellison}}, \bibinfo {author}
  {\bibfnamefont {J.~X.}\ \bibnamefont {Prochaska}}, \bibinfo {author}
  {\bibfnamefont {L.}~\bibnamefont {Kaper}}, \bibinfo {author} {\bibfnamefont
  {J.}~\bibnamefont {Tumlinson}}, \bibinfo {author} {\bibfnamefont {R.~F.}\
  \bibnamefont {Carswell}}, \ and\ \bibinfo {author} {\bibfnamefont
  {W.}~\bibnamefont {Ubachs}},\ }\href {\doibase
  10.1111/j.1365-2966.2009.16227.x} {\bibfield  {journal} {\bibinfo  {journal}
  {MNRAS}\ }\textbf {\bibinfo {volume} {403}},\ \bibinfo {pages} {1541}
  (\bibinfo {year} {2010})}\BibitemShut {NoStop}%
\bibitem [{\citenamefont {van Weerdenburg}\ \emph {et~al.}(2011)\citenamefont
  {van Weerdenburg}, \citenamefont {Murphy}, \citenamefont {Malec},
  \citenamefont {Kaper},\ and\ \citenamefont {Ubachs}}]{Weerdenburg2011}%
  \BibitemOpen
  \bibfield  {author} {\bibinfo {author} {\bibfnamefont {F.}~\bibnamefont {van
  Weerdenburg}}, \bibinfo {author} {\bibfnamefont {M.~T.}\ \bibnamefont
  {Murphy}}, \bibinfo {author} {\bibfnamefont {A.~L.}\ \bibnamefont {Malec}},
  \bibinfo {author} {\bibfnamefont {L.}~\bibnamefont {Kaper}}, \ and\ \bibinfo
  {author} {\bibfnamefont {W.}~\bibnamefont {Ubachs}},\ }\href {\doibase
  10.1103/PhysRevLett.106.180802} {\bibfield  {journal} {\bibinfo  {journal}
  {Phys. Rev. Lett.}\ }\textbf {\bibinfo {volume} {106}},\ \bibinfo {pages}
  {180802} (\bibinfo {year} {2011})}\BibitemShut {NoStop}%
\bibitem [{\citenamefont {van Veldhoven}\ \emph {et~al.}(2004)\citenamefont
  {van Veldhoven}, \citenamefont {K\"{u}pper}, \citenamefont {Bethlem},
  \citenamefont {Sartakov}, \citenamefont {van Roij},\ and\ \citenamefont
  {Meijer}}]{Veldhoven2004}%
  \BibitemOpen
  \bibfield  {author} {\bibinfo {author} {\bibfnamefont {J.}~\bibnamefont {van
  Veldhoven}}, \bibinfo {author} {\bibfnamefont {J.}~\bibnamefont
  {K\"{u}pper}}, \bibinfo {author} {\bibfnamefont {H.~L.}\ \bibnamefont
  {Bethlem}}, \bibinfo {author} {\bibfnamefont {B.}~\bibnamefont {Sartakov}},
  \bibinfo {author} {\bibfnamefont {A.~J.~A.}\ \bibnamefont {van Roij}}, \ and\
  \bibinfo {author} {\bibfnamefont {G.}~\bibnamefont {Meijer}},\ }\href
  {http://dx.doi.org/10.1140/epjd/e2004-00160-9} {\bibfield  {journal}
  {\bibinfo  {journal} {Eur. Phys. J. D}\ }\textbf {\bibinfo {volume} {31}},\
  \bibinfo {pages} {337} (\bibinfo {year} {2004})}\BibitemShut {NoStop}%
\bibitem [{\citenamefont {Flambaum}\ and\ \citenamefont
  {Kozlov}(2007)}]{FlambaumNH3_2007}%
  \BibitemOpen
  \bibfield  {author} {\bibinfo {author} {\bibfnamefont {V.~V.}\ \bibnamefont
  {Flambaum}}\ and\ \bibinfo {author} {\bibfnamefont {M.~G.}\ \bibnamefont
  {Kozlov}},\ }\href {\doibase 10.1103/PhysRevLett.98.240801} {\bibfield
  {journal} {\bibinfo  {journal} {Phys. Rev. Lett.}\ }\textbf {\bibinfo
  {volume} {98}},\ \bibinfo {pages} {240801} (\bibinfo {year}
  {2007})}\BibitemShut {NoStop}%
\bibitem [{\citenamefont {Murphy}\ \emph {et~al.}(2008)\citenamefont {Murphy},
  \citenamefont {Flambaum}, \citenamefont {Muller},\ and\ \citenamefont
  {Henkel}}]{Murphy2008}%
  \BibitemOpen
  \bibfield  {author} {\bibinfo {author} {\bibfnamefont {M.~T.}\ \bibnamefont
  {Murphy}}, \bibinfo {author} {\bibfnamefont {V.~V.}\ \bibnamefont
  {Flambaum}}, \bibinfo {author} {\bibfnamefont {S.}~\bibnamefont {Muller}}, \
  and\ \bibinfo {author} {\bibfnamefont {C.}~\bibnamefont {Henkel}},\ }\href
  {\doibase 10.1126/science.1156352} {\bibfield  {journal} {\bibinfo  {journal}
  {Science}\ }\textbf {\bibinfo {volume} {320}},\ \bibinfo {pages} {1611}
  (\bibinfo {year} {2008})}\BibitemShut {NoStop}%
\bibitem [{\citenamefont {Kanekar}(2011)}]{Kanekar2011}%
  \BibitemOpen
  \bibfield  {author} {\bibinfo {author} {\bibfnamefont {N.}~\bibnamefont
  {Kanekar}},\ }\href {http://stacks.iop.org/2041-8205/728/i=1/a=L12}
  {\bibfield  {journal} {\bibinfo  {journal} {Astroph. J. Lett.}\ }\textbf
  {\bibinfo {volume} {728}},\ \bibinfo {pages} {L12} (\bibinfo {year}
  {2011})}\BibitemShut {NoStop}%
\end{thebibliography}
\end{document}